\documentclass[journal,draftclsnofoot,onecolumn,twoside,12pt]{IEEEtran}

\usepackage[T1]{fontenc}
\usepackage{amsmath} 
\usepackage{amssymb}
\usepackage{algorithmic}
\usepackage{cite}
\usepackage{dsfont} 
\ifCLASSINFOpdf
   \usepackage[pdftex]{graphicx}
\else
  \usepackage[dvips]{graphicx}
\fi

\newtheorem{property}{Property}
\newtheorem{theorem}{Theorem}
\newtheorem{proposition}{Proposition}
\newtheorem{remark}{Rmark}

\DeclareMathOperator*{\argmin}{argmin}

\begin{document}
\title{Modulation-Specific Multiuser Transmit Precoding and User Selection for BPSK Signalling}

\author{Majid Bavand, \IEEEmembership{Student Member, IEEE,} Steven D. Blostein,
\IEEEmembership{Senior Member, IEEE}
\thanks{Part of this work was presented at 49th Asilomar Conference on
Signals, Systems, and  Computers, Pacific Grove, CA, November 2015.

The authors are with the Department
of Electrical and Computer Engineering, Queen's University, Ontario,
Canada, K7L 3N6 (e-mail: \{m.bavand, steven.blostein\}@queensu.ca).}
}
\maketitle

\begin{abstract}
Motivated by challenges to existing multiuser transmission methods in a
low signal to noise ratio (SNR) regime,
and emergence of massive numbers of low data rate ehealth and internet of things (IoT)
devices, in this paper we show that it is beneficial
to incorporate knowledge of modulation type into multiuser
transmit precoder design. Particularly, we propose a transmit precoding (beamforming)
specific to BPSK modulation, which has maximum power efficiency and capacity in poor
channel conditions.
To be more specific, in a multiuser scenario, an objective function is formulated based on
the weighted sum of error probabilities of BPSK modulated users.
Convex optimization is used to transform and solve this ill-behaved non-convex
minimum probability of error (MPE) precoding problem.
Numerical results confirm significant performance improvement.
We then develop a low-complexity user selection algorithm for MPE precoding.
Based on line packing principles in Grassmannian manifolds, the
number of supported users is able to exceed the number of transmit
antennas, and hence the proposed approach is able to support more simultaneous users
compared with existing multiuser transmit precoding methods.
\end{abstract}

\section{Introduction}\label{secInro_bpsk}
\IEEEPARstart{W}{ireless} multiple input multiple output (MIMO) channels have been
attracting a great deal of interest in the last decades
\cite{Valenzuela98,Schubert04,Peel05,Goldsmith06}.
MIMO technologies, at the core of several wireless standards,
improve spectral efficiency and reliability compared to
single-input single-output (SISO) systems.
MIMO system design has been usually posed under two different perspectives: either
to increase data transmission rate through spatial multiplexing or to improvement
system reliability through increased antenna diversity.
Spatial multiplexing is a simple MIMO transmit technique that it does not require channel
state information (CSI) at the transmitter and enables high spectral efficiency by
splitting the incoming data into
multiple independent substreams and transmitting each substream on a different antenna
as in V-BLAST \cite{Valenzuela98}.
When CSI is available at the transmitter, channel-dependent linear or nonlinear transmit precoding
(beamforming) of the data substreams
can further improve the performance by adapting the transmitted signal
to the instantaneous channel state \cite{Miguel09}.
In this case, employing multiuser MIMO techniques
allows for a gain in sum capacity obtained by channel
reuse \cite{Schubert04,Peel05,Goldsmith06,Sadek07WC}.

Although channel reuse for multiple users is advantageous in terms of throughput, when
multiple uncoordinated links share a common communications medium,
e.g., in a broadcast system,
co-channel interference caused by the transmission of multiple users' data on the
same carrier frequency could limit channel reuse \cite{Etkin08}.
Most wireless systems avoid interference by orthogonalizing the communication
links in time or frequency.
It is clear that this approach could be suboptimal since it entails a priori loss of
degrees of freedom in both links independent of the amount of interference.
Power control, precoding, and scheduling techniques, with capability of reducing
interference, are conventional solutions to the co-channel interference problem
\cite{Goldsmith06}.
From a practical point of view, using multiple antennas to communicate
with many users simultaneously is especially appealing in
wireless local area network (WLAN) environments, WiMAX,  and other time-division duplex (TDD)
systems where channel conditions can readily be learned by all parties \cite{Peel05}.

Classically, a beamformer or precoder controls the beam pattern of an antenna array by
weighting antennas to satisfy predetermined optimization criteria.
Many precoding methods aim at maximizing throughput.
However, as pointed out by Palomar et al.
\cite{Palomar03}, the problem with this type of criterion is that it implicitly
presumes that an unrealizable ideal continuous Gaussian code is used instead of
a signal constellation.
In practice, the transmitter sends a modulated signal with
a practical suboptimal channel coding scheme which together
determine system throughput.
Signal to noise ratio (SNR), signal to interference plus noise ratio (SINR),
mean square error (MSE) between the desired signal and the array output,
and signal to leakage ratio (SLR)
are other common criteria in formulating the precoding optimization problem
\cite{Schubert04,Peel05,Sadek07WC}.
However, in the communications scenario considered here, the probability of
error or achievable bit error rate (BER) is the system performance metric that
maximizes capacity for BPSK transmission
\cite{MajidAsilomar15,Alouini13,Hanzo05,BlosteinVT07,Antoniou00}.
Therefore, designing the transmit precoder to directly minimize the error probability
would result in improved system performance.

The error probability of each user in a multiuser downlink system depends on
the modulation type.
Therefore, to design a precoder that minimizes
error probability, one ought to account for the modulation type which is
not considered in classical precoding methods such as
minimum mean square error (MMSE), maximum signal to leakage and noise ratio (MSLNR),
and block diagonalization (BD) \cite{Spencer04}. 
In this paper, it is established that by incorporating modulation type in the
precoder design, system performance may be significantly improved.

As additional motivation, low data rate BPSK modulation is a commonly
employed transmission mode in adaptive wireless systems such as IEEE 802.11a,n,ac,
when SNR is low \cite{Saeed14,Forney98}.
Moreover, emerging technologies such as internet of things (IoT) require
simultaneous deployment of a massive number of low data rate devices,
which serves as another motivation for employing BPSK modulation \cite{Nokia15}.
Finally, in current WiFi systems such as 802.11ac, the transmit precoding is
considered as a selectable adaptive ``MIMO mode'' in addition to modulation type,
which naturally motivates coupling of transmit precoding to modulation.

As mentioned earlier, selecting a subset
of users for transmission in a broadcast channel, is another
conventional method to reduce co-channel interference
and increase system throughput and reliability.
Gains in throughput and reliability are also obtained by multiuser diversity
via user selection when the number of users is large.
Although the optimal user subset can be found by brute-force search over all
possible combinations of user subsets, its computational complexity
is prohibitive. In practice, low-complexity scheduling algorithms are desired
\cite{Goldsmith06,Mao12,Evans06,Jindal12,Ammar15,Leung13}.
For example in \cite{Goldsmith06,Mao12}, algorithms based on
semi-orthogonal user selection (SUS) are presented, which are developed for
zero-forcing precoding. When SUS is combined with zero-forcing precoding and
water filling power allocation, although overall suboptimal, it can achieve
the same asymptotic (high SNR) sum rate as that of dirty paper coding
for broadcast channel, as the number of users goes to infinity.
In \cite{Evans06}, a greedy user selection algorithm
is proposed based on BD precoding and increases total throughput
of users. In this paper, we propose a user selection algorithm
for MPE precoding that semi-greedily selects the set of users
by a geometric approach such that the number of selected users is made as large
as possible. It is shown that for a one-dimensional modulation such as
binary phase-shift keying (BPSK), it is possible to utilize extra dimensions
provided by the complex channel and transmit information at the same time and
frequency to more users than the number of transmit antennas.

The contributions of this paper are summarized as follows:
\begin{enumerate}
  \item A new multiuser transmit precoder that minimizes the probability of error
  is proposed for BPSK signalling in the multiuser multiple input single output
  (MISO) broadcast channel for the following scenarios:
  (i) single-user maximum likelihood detection at the receivers, and (ii) joint
  transmit precoding and receive filtering.
  \item A low-complexity geometric user selection (GUS) algorithm is developed for MPE transmit precoding,
  which can select more simultaneous users than the number of transmit antennas.
\end{enumerate}

The rest of this paper is organized as follows: in Section
\ref{secSysModel_bpsk}, the system model is introduced. In Section \ref{secPe_bpsk},
the error probability of a user in the downlink of a multiuser system is calculated
assuming the transmitter is using BPSK modulation and linear precoding for
transmission. Section \ref{secMpe_bpsk} presents the minimum probability of
error (MPE) transmit precoding considering two different scenarios.
First, it is assumed that the receiver uses a single-user maximum likelihood detector.
Second, it is assumed that the transmit precoding weights and the receive filter
coefficients are calculated jointly at the transmitter.
In both scenarios we try to transform the precoding optimization
problem to convex optimization subproblems and present algorithms for finding the precoding
vectors of users. In Section \ref{secUserSelection_bpsk}, by taking a geometric approach,
a user selection algorithm compatible
with MPE precoding is presented. Numerical results are demonstrated
in Section \ref{secResults_bpsk}. Finally, conclusions are drawn in
Section \ref{secConclusion_bpsk}.

\section{System Model}\label{secSysModel_bpsk}
\begin{figure}[tp]
  \centering
  \includegraphics[width=3.49in]{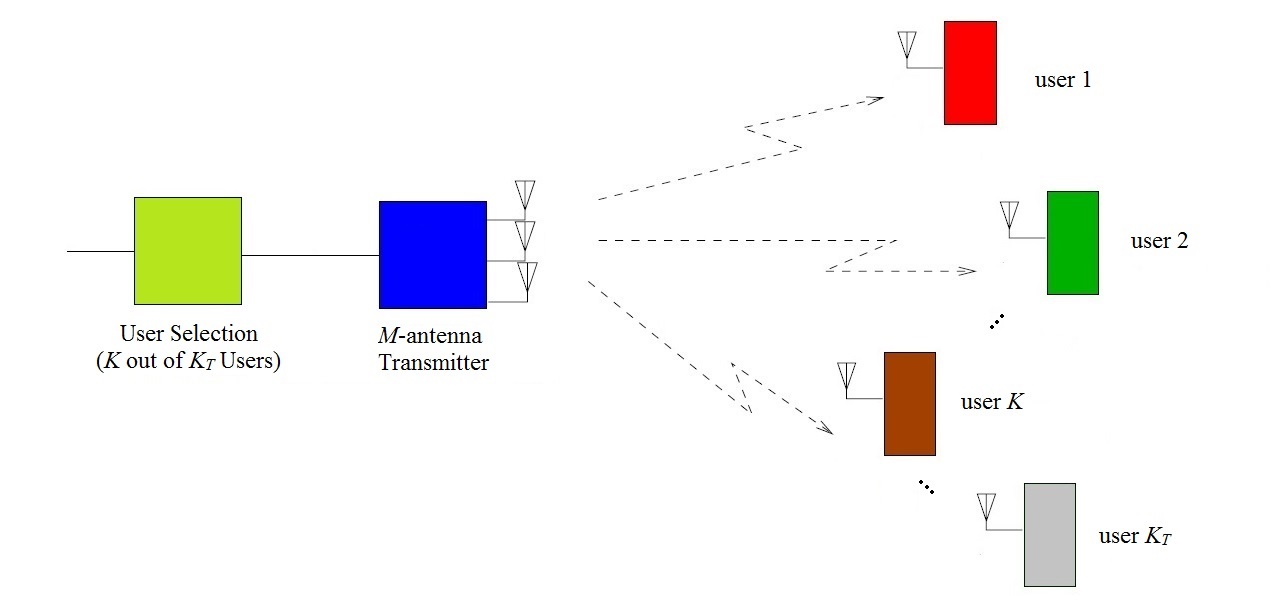}\\ 
  \caption{System model.} 
  \label{figure1}
\end{figure}

We consider a broadcast system with one transmitter
and $K$ receivers (users). Later in Section \ref{secUserSelection_bpsk}, it is
shown that these $K$ users are preselected out of $K_T$ total available users by the
proposed algorithm in Section \ref{secUserSelection_bpsk}. The system model
is shown in Fig. \ref{figure1}. It is assumed that the transmitter
consists of an array of antennas with $M$ elements and each receiver
$j,~1 \leq j \leq K$, has one antenna in
its array. It is also assumed that the transmitter has one symbol encoded
in ${s}_j$ for each receiver $j$, $1 \leq j \leq K$ to be transmitted in
the same time and frequency slots.
The transmitter uses an $M \times 1$ precoding vector ${\bf u}_j$ to
encode the transmitted symbols intended for receiver $j$.
Ignoring the noise at the transmitter's output, and assuming that the transmitter
has information symbols for all of the $K$ preselected receivers we can model the
$M \times 1$ transmitted signal vector as
\begin{equation*}\label{x_eq1}
    {\bf x} = \sum_{l=1}^{K}{\bf u}_{l} {s}_{l} = \bf U \bf s,
\end{equation*}
where ${\bf U} = [{\bf u}_1,~\cdots,~{\bf u}_K]$ and ${\bf s} = [{s}_1, \ldots, {
s}_K]^T$. The channel matrix between $M$ antennas of the transmitter and the single antenna
of receiver $j$ can be represented by the $1 \times M$ vector ${\bf h}_{j}$
with entries following an independent identically distributed (i.i.d.) circularly
symmetric complex Gaussian (CSCG) distribution with zero mean and unit variance.
This channel model is valid for narrow-band (frequency non-selective) systems if the
transmit and receive antennas are in
non line-of-sight rich-scattering environments with sufficient antenna spacing \cite{PaulrajBook03,Mao12}.
It should be remarked that we follow the same vector representation of the channel
as \cite{Goldsmith06}, i.e., representing the channels with row vectors in MISO systems.
At receiver $j$ the received signal can be modeled as
\begin{equation*}\label{r_eq2}
    {r}_j = {{\bf h}_{j} {\bf x} + {z}_j} =
    {\sum_{l=1}^{K}{{\bf h}_{j} {\bf u}_{l} {s}_{l}}}
    + {z}_j = \bar{{r}}_j+ {z}_j, \quad 1 \leq j \leq K,
\end{equation*}
where ${z}_j$ is a CSCG noise with zero mean and
variance $\sigma _z^2$.

At receiver $j$ the received signal is processed by a filter.
Therefore the output at the receiver $j$ can be written as
\begin{align*}\label{y_eq3}
    {y}_{j} = y_j(w_j,{\bf U}) = {w}_{j} {r}_j = {\sum_{l=1}^{K}
    {{w}_{j} {\bf h}_{j} {\bf u}_{l} {s}_{l}}}
    + {w}_{j}{z}_j = \bar{{y}}_{j}+ {z}'_{j},\\
    \quad 1 \leq j \leq K,
\end{align*}
where ${w}_{j}$ is the complex-valued filter coefficient
at receiver $j$ and ${z}'_{j}={w}_{j}{z}_j$ is complex Gaussian noise
with variance $\sigma^2_z {w}_{j} {w}_{j}^H$. We also express $y_j$ as a function
of $w_j$ and ${\bf U}$ to emphasize its dependence on these parameters.

\section{Error Probability}\label{secPe_bpsk}
To calculate the error probability of each user in a broadcast channel one needs to
know the estimation technique that the receiver is using in addition to the
modulation type. 
Motivated by the facts that a low dimensional modulation like BPSK is
employed in wireless systems such as
IEEE 802.11a,n,ac, and it is often selected in systems with adaptive
modulation when SNR is low, we assume BPSK modulation for all users \cite{Saeed14,Forney98}.
It is also assumed that the system operates with the following decision rule for estimating the
transmitted symbols of user $j$ when the output noise is additive white Gaussian:
\begin{equation*}\label{eq6}
    {\hat s}_j = {\rm sign}(y_j^R),
\end{equation*}
where the superscript $^R$ denotes taking the real part operation, i.e., $x^R={\rm {Re}}\{x\}$.

Here, the error probability of each user $j,~1 \leq j \leq K,$ is
calculated as a function of its receive filter coefficient and its
transmit precoding weight vector as well as the transmit precoding
weight vectors of all other users
($P_{e_j}(w_j;{\bf U}),~1 \leq j \leq K$).
Later in Section \ref{secMpe_bpsk}, we use this probability of error to calculate
precoding weights for two different scenarios, namely,
a scenario in which each receiver estimates the transmitted signal according to a
classical single-user maximum likelihood scheme without relying on knowledge of
other channels, and a scenario with joint transmit precoding and receive filtering in which
both of the receive filtering and transmit precoding weights are calculated at the transmitter
and then the calculated receive filter coefficient of each user is provided to
its receiver via the forward channel.

The error probability at the output of receiver $j$ is expressed as
\begin{align*}
    \nonumber
    P_{e_j} & = P_{e_j}(y_j;w_j,{\bf U}) = 
    P(\hat{s}_j \neq s_j) \\ \nonumber
    & = P_0 P({\hat s}_j = 1 | s_j = -1) + P_1 P({\hat s}_j = -1 | s_j = 1) \\
    & = P(\hat{s}_j=-1|s_j=1) = P(y_j^R < 0|s_j=1),
\end{align*}
where $P_0 = P(s_j = -1)$, $P_1 = P(s_j = +1)$, and $P(.)$ is the
probability of an event. It is assumed that $P_0 = P_1 = 1/2$,
i.e., the transmitted BPSK modulated signal $s_j$ takes its elements from the
set $\{\pm 1\}$ with equal probability,
which could be the result of source compression and hard decision.
It should be remarked that, using uniform rather than Gaussian
distribution over signal sets causes an asymptotic loss in throughput
which could be compensated to some extent by using constellation shaping
techniques \cite{Forney98}.
The symmetry in the estimation of $+1$ and $-1$
for user $j$ is also considered, which is the result of the symmetry
between $p(y_j^R| s_j = -1)$ and
$p(y_j^R | s_j = +1)$, where $p$ is the probability density function (pdf),
and consequently results in $P({\hat s}_j = -1 | s_j = +1)=P({\hat s}_j = +1 | s_j = -1)$.
Since the number of constellation points in BPSK is two,
$N_{b}=2^K$ is the number of possible symbol sequences for
all $K$ users in one transmission, i.e., there could be $2^K$ different possible sets
of $K$-tuple symbols ${\bf s}_b,~ 1 \leq b \leq N_b,$ for $K$ users.
For BPSK transmission, we denote
$N_{pb}=2^{K-1}$ as the number of possible symbol sequences for transmission if the
transmitted symbol of user $j$ is already known for example to be $+1$.

Using equal probability for transmission of BPSK
constellation points, and additive Gaussian output noise $\rm{Re}\{z'_{j}\}$, we have
\begin{align*}
    &p(y_j^R | s_{j} = +1)\\
    &= \frac{1}{N_{pb}}
    \sum_{s_{l\ne j} \in \{\pm 1\} }
    p(y_j^R | s_1, \ldots, s_j = +1, s_{j+1}, \ldots, s_K) \\
    &=
    \frac{1}{N_{pb}}
    \sum_{\substack{ b = 1 \\ s_j \neq -1}}^{N_{pb}}
    \frac{1}{\sqrt{\pi \sigma_z^2 w_j^H w_j}} \exp
    {-\frac{(y_j^R-\bar{y}_{j,b}^R(w_j,{\bf U}))^2}
    {\sigma_z^2 |w_j|^2}},
\end{align*}
where in the first equality, using total probability theorem, the conditional output
probability at receiver $j$ is again conditioned over all $2^{K-1}$ possible
assignments of transmitted symbols $s_{l \ne j}$.
Also $\bar{y}_{j,b}^R(w_j,{\bf U})$ is ${\rm Re}\{\bar{y}_{j}\}$ when
${\bf s}_b$ is transmitted, i.e.,
\begin{equation*}
    \bar{y}_{j,b}^R(w_j,{\bf U}) = {\rm{Re}}\{w_j {\bf h}_j {\bf U} {\bf s}_b \},
    \quad 1 \leq b \leq N_b.
\end{equation*}
Therefore, the probability of error can be calculated as
\begin{align} \label{pej_joint}
    \nonumber
    P_{e_j} \!&=\! \int_{-\infty}^0 {\!p(y_j^R | s_j \!=\! +1) dy_j^R} \!=\!
    \frac{1}{N_{pb}}
    \!\sum_{\substack{b = 1 \\ s_j \neq -1}}^{N_{pb}}
    {\!\!Q\!\left(\!\frac{\bar{y}_{j,b}^R}{\sqrt{\frac{\sigma_z^2 | w_j|^2}{2}}}\!\right)\!} \\
    &=\frac{1}{N_{b}}
    \sum_{b = 1}^{N_b}
    {Q\left(\frac{s_{b,j}{\rm{Re}}\{w_j {\bf h}_j \sum_{l=1}^K{
    {\bf u}_l {s}_{b,l}} \}} {\frac{\sigma_z}{\sqrt{2}} |w_j|}\right)},
\end{align}
where $s_{b,j}$ is the $j$th symbol of ${\bf s}_b$ and
$Q(x)$ is defined as $\frac{1}{\sqrt{2\pi}} 
\int_x^{\infty} {e^{-\frac{u^2}{2}}du}$.

\section{Minimum Probability of Error Precoding}\label{secMpe_bpsk} 
In this section, our objective is to minimize the weighted sum of error probabilities
for two different scenarios:
\begin{enumerate}
  \item First scenario- the receiver only needs its own
  channel information to calculate its filter coefficient and there is no need for
  extra feedback. It is assumed that the receivers use ML detection. We try to
  minimize the weighted sum of
  error probabilities only over transmit precoding weights since each receive filter
  coefficient could be expressed as a function
  of transmit precoding weights. This method, while yielding suboptimal performance
  as compared to joint transmit precoding-receive filtering is nevertheless easier to
  practically realize.
  \item Second scenario- we assume joint transmit precoding-receive filtering in which
  we minimize the error sum over both transmit precoding and receive filtering weights. This
  approach could be considered as the optimal method for communications in
  broadcast system when considering the error probability as the measure of quality.
\end{enumerate}
If one wants to minimize error probabilities of all users, it means
that several objective functions have to be minimized which are all interdependent
by the common transmit precoding matrix.
A standard approach to this multiple-objective optimization problem
is to combine the individual objective functions into a single composite function
\cite{Konak06,Bjornson14}.
Similar to \cite{Palomar03}, we use the weighted average error probability of users
$    P_e = \frac{1}{\sum_{j=1}^K \alpha_j} \sum_{j=1}^K \alpha_j P_{e_j}$.
If $\alpha_j = 1$ for all users, i.e., all users have the same priority, the weighted
average turns to a simple average of the error probability of users in (\ref{pej_joint}) as
\begin{align}\label{avg_pe}
    \nonumber
    P_e &= \frac{1}{K} \sum_{j=1}^K P_{e_j} \\
    & = \frac{1}{KN_b} \sum_{j=1}^K
    \sum_{b = 1}^{N_b}
    {Q\left(\frac{s_{b,j}{\rm{Re}}\{w_j {\bf h}_j\sum_{l=1}^K{
    {\bf u}_l {s}_{b,l}} \}} {\frac{\sigma_z}{\sqrt{2}} |w_j|}\right)},
\end{align}
which is considerd as the system performance criterion hereafter.
The parameters ${\bf w} = [w_1,\ldots,w_K]$ and ${\bf U}$ have a
bilinear dependence in the numerator of each $Q$-function argument of (\ref{avg_pe}).
Now, we state the following constraints that are required in both forthcoming
scenarios.
\begin{proposition} \label{lem_positive}
    For the $j$th user not to have an error floor (a rough lower bound on
    error probability), it is necessary
    for $w_j$ to comply with the following constraints:
    \begin{equation}\label{eq_posit_constr}
        {s_{b,j}{\rm{Re}}\left\{w_j \sum_{l=1}^K{{\bf h}_j
        {\bf u}_l {s}_{b,l}} \right\}} \geq 0,\quad~~ 1 \leq b \leq N_b.
    \end{equation}
\end{proposition}
\begin{IEEEproof}
    See Appendix \ref{append_lem_positive}.
\end{IEEEproof}

\subsection{Transmit Precoding with ML Receiver}\label{subSecMpeML_bpsk}
For the first scenario, it is assumed that the receivers use single-user maximum
likelihood detection similar to \cite{Sadek07WC}. Additionally, for the
transmit precoding design the receive filtering weights are assumed to be
available at the transmitter, because they could be calculated in closed-form at
the transmitter.
For ML detection the receive filter coefficient of a user can be expressed
as a function of its transmit precoding vector and channel as \cite{Sadek07WC}
\begin{equation*}
    w_j = \frac{{\bf u}_j^H{\bf h}_j^H}{|{\bf h}_j{\bf u}_j|^2}, \quad 1 \leq j \leq K.
\end{equation*}
Consequently, the error probability of a user in (\ref{pej_joint})
will be only a function of ${\bf U}$ and so will be the average error probability of
users in (\ref{avg_pe}):
\begin{equation}\label{pe_ML}
    P_{e}^{\text{ML}}({\bf U}) =
    \frac{1}{KN_{b}}
    \sum_{j = 1}^{K}\sum_{b = 1}^{N_b}
    {Q\!\left(\!\frac{\sum_{l=1}^K{\rm{Re}\{{\bf u}_j^H {\bf h}_j^H {\bf h}_j
    {\bf u}_l {s}_{b,l}s_{b,j} \}}}{\frac{\sigma_z}{\sqrt{2}}|{\bf h}_j
    {\bf u}_j|} \!\right)\!}.
\end{equation}
Now, we try to minimize the average error probability $P_{e}^{\text{ML}}({\bf U})$
by constraining the total transmit power.
Based on Proposition \ref{lem_positive},
the arguments of $Q$-functions are constrained to be
nonnegative. Thus, the transmit precoding optimization problem is stated as
\begin{align*}\label{ML_min}
    \nonumber
    &\min_{\bf U} P_e^{\text{ML}}  \\
    \nonumber
    &{\text{subject to}} \quad {\rm Tr}({\bf U}{\bf U}^H) \leq \tau,\\
    & \sum_{l=1}^K{{\rm Re}\{ {\bf u}_j^H {\bf h}_j^H
    {\bf h}_j {\bf u}_l s_{b,l}s_{b,j}  \}} \geq 0 ,~ 1 \leq j \leq K, ~1 \leq b \leq N_b,
\end{align*}
where $\tau$ is the total transmit power constraint.

Now, let us restate the transmit precoding vector ${\bf u}_j$ of each user as
\begin{equation*}
    {\bf u}_j = a_j {\bf \bar u}_j,\quad 1\leq j \leq K,
\end{equation*}
where $a_j = \|{\bf u}_j\|_2$ is the amplitude of ${\bf u}_j$ and
${\bf\bar u}_j = \frac{{\bf u}_j}{\|{\bf u}_j\|_2}$, i.e., $\|{\bf \bar u}_j\|_{2}=1$.
By using the Cauchy-Schwarz inequality the following upper bound on the error
probability (\ref{pe_ML}) is obtained:
\begin{equation}\label{pe_ML_ub}
    P_{e}^{\text{ML-Up}} =
    \frac{1}{KN_{b}}
    \sum_{j = 1}^{K}\sum_{b = 1}^{N_b}
    {Q(\frac{\sum_{l=1}^K{{\rm Re}\{{\bf \bar u}_j^H {\bf h}_j^H {\bf h}_j
    {\bf \bar u}_l {s}_{b,l}s_{b,j} a_l \}}}{\frac{\sigma_z}{\sqrt{2}}\|{\bf h}_j\|} )}.
\end{equation}
The Cauchy-Schwarz inequality and therefore the upper bound is tight if ${\bf u}_j = {\bf h}_j^H$,
which can occur when all channels are orthogonal to each other.
Now, we minimize (\ref{pe_ML_ub}) as follows:
\begin{subequations}\label{ML_min_mod1}
\begin{align}
    & \min_{{\bf \bar U},{\bf a}}P_{e}^{\text{ML-Up}} \label{ML_min_mod1_obj} \\
    & {\text {subject to}} \quad \sum_{j=1}^K a_j^2 \leq \tau,
    \label{ML_min_mod1_first_const}\\
    &|{\bf h}_j{\bf\bar u}_j|^2 a_j \!\!+\!\! \sum_{\substack{l=1 \\ l \ne j}}^K
    \!{{\rm Re}\{ {\bf \bar u}_j^H {\bf h}_j^H
    {\bf h}_j {\bf \bar u}_l s_{b,l}s_{b,j} a_l  \}} \!\geq\! 0,~
    \begin{array}{l}  \\
    1 \!\leq\! j \!\leq\! K,\\
    1 \!\leq\! b \!\leq\! N_b,
    \end{array}\!\!\!\!
    \label{ML_min_mod1_second_const} \\
    &\|{\bf \bar u}_j \| = 1, \quad 1 \le j \le K, \label{ML_min_mod1_third_const}
\end{align}
\end{subequations}
where ${\bf \bar U} = [{\bf\bar u}_1,\cdots, {\bf\bar u}_K ]$
and  ${\bf a} = [a_1,\cdots,a_K]$.

We propose the alternating minimization algorithm of Table \ref{table_ML}
which minimizes $P_e^{\text{ML-U}}$ over ${\bf\bar U}$ and ${\bf a}$ alternatingly
to solve (\ref{ML_min_mod1}).
\begin{table}[tp]
    \centering
    \caption{Alternating minimization algorithm for solving (\ref{ML_min_mod1})}
    \label{table_ML}
    \begin{tabular}{l}
        \hline\hline
        \hspace{-.13in}\begin{minipage}[]{3.45in}
            \begin{algorithmic}
                \STATE
                \STATE {\bf Initialization:} \\
                \STATE ${\bf\bar U^0} \gets$ random complex $M \times K$ matrix such that each
                column is normalized.\\
                \STATE ${\bf a^0} \gets$ random real positive $1 \times K$ vector such that
                $\sum_{j=1}^K ({a^0_j})^2 \leq \tau$.\\
                \STATE Initialize ${P_e^1}$ and $P_e^2 < P_e^1$ with
                proper values to start the while loop, e.g. $P_e^1 = 1$
                and $P_e^2 = 0.5 $.\\
                \WHILE  {$P_e^1-P_e^2 > P_e^{\text{threshold}}$}
                \STATE ${P_e^1} \gets {P_e^2}$.\\
                \STATE {\bf \underline{Minimization 1:}}
                \STATE Minimize (\ref{ML_min_mod1}) over ${\bf\bar U}$
                assuming that ${\bf a} = {\bf a^0}$, and using the initial
                value of ${\bf\bar U^0}$ for ${\bf\bar U}$.\\
                \STATE ${\bf\bar U^0} \gets {\bf\bar U}^{\text{opt}}$.\\
                \STATE {\bf \underline{Minimization 2:}}
                \STATE Minimize (\ref{ML_min_mod1}) over ${\bf a}$ assuming
                ${\bf\bar U} = {\bf\bar U^0}$, by using a numerical convex optimization method
                and the initial value of  ${\bf a^0}$ for ${\bf a}$.
                \STATE ${\bf a^0} \gets {\bf a}^{\text{opt}}$.\\
                \STATE ${P_e^2} \gets P_e^{\text{ML-U}}({\bf\bar U^0},{\bf a^0})$.\\
                \ENDWHILE
                \STATE ${\bf U} = {\bf\bar U^0}{\rm diag}({\bf a})$ is the transmit precoding matrix.\\
            \end{algorithmic}
        \end{minipage}
        \\
        \hline\hline
    \end{tabular}
\end{table}
The algorithm iterates until the average error probability converges with the accuracy
of $P_e^{\text{threshold}}$.
In Minimization 1 of the proposed algorithm, when (\ref{ML_min_mod1_obj})
is minimized over ${\bf\bar U}$, constraint (\ref{ML_min_mod1_first_const})
does not depend on the minimization parameter and can be removed.
Similarly, in Minimization 2, when (\ref{ML_min_mod1_obj})
is minimized over ${\bf a}$, constraints in (\ref{ML_min_mod1_third_const})
are not included since they are independent of the optimization parameters.
Now, we observe that the alternating minimization algorithm has the
following properties:

\begin{property}\label{Proper_convex_ML} 
    Minimization 2 in Table \ref{table_ML} is a convex optimization problem.
\end{property}
\begin{IEEEproof}
    See Appendix \ref{append_Proper_convex_ML}
\end{IEEEproof}

\begin{theorem} \label{Theo_converge_ML}
    The algorithm presented in Table \ref{table_ML} converges to a local minimum of
    (\ref{ML_min_mod1}).
\end{theorem}
\begin{IEEEproof}
    The objective function $P_e^{\text{ML-U}}({\bf\bar U},{\bf a})$
    is minimized in Minimization 1 of Table \ref{table_ML}
    over ${\bf\bar U}$ while keeping ${\bf a}$ fixed at the values obtained in Minimization
    2 of the previous iteration. Therefore, at each iteration of the algorithm the value
    of the objective function after
    Minimization 1 is non-increasing compared to the value of the objective function
    after Minimization 2 of the previous iteration.
    In Minimization 2 of Table \ref{table_ML}, $P_e^{\text{ML-U}}({\bf\bar U},{\bf a})$
    is minimized over ${\bf a}$ while keeping ${\bf\bar U}$ at the values obtained
    in Minimization 1. Therefore, the value of $P_e^{\text{ML-U}}$ does not increase
    compared to the result of Minimization 1 of the same iteration.
    Therefore, each iteration of the algorithm causes $P_e^{\text{ML-U}}$ to be
    non-increasing. Moreover, considering the fact that
    $P_e^{\text{ML-U}}$ is always nonnegative, $0 \leq P_e^{\text{ML-U}} \leq 1$, guarantees
    that the algorithm converges to a stationary (minimum) point.
\end{IEEEproof}

\subsection{Joint Transmit Precoding-Receive Filtering}\label{subSecMpeJoint_bpsk}
For the second scenario, it is assumed that the receive filter coefficients
are calculated by the transmitter in conjunction with the transmit precoding
weights, and each receiver is provided with its receive filter coefficient.
In this scenario, the transmitter finds the transmit precoding weights and receive
filter coefficients that minimize the average error probability of users.
Therefore, the joint MPE transmit precoding-receive filtering problem is represented by
\begin{align} \label{joint_min}
    \nonumber
    & \min_{{\bf w}, {\bf U}} P_e \\
    & {\text {subject to}} \quad {\rm Tr}({\bf U}{\bf U}^H) \leq \tau,
\end{align}
where $P_e$ is the average error probability of (\ref{avg_pe}).
It should be noted that the error probability of user $j$, $P_{e_j}$ in (\ref{pej_joint}),
depends on its receive filter coefficient $w_j$, and the transmit precoding matrix
of all users ${\bf U}$, but it does not directly depend on the
receive filter coefficients of other users\footnote{Although explicitly $P_{e_j}$
does not depend on $w_{k\ne j}$, it indirectly is coupled with them, because ${\bf U}$
depends on all receive filters.}.
Based on this observation
we strive to develop an alternating minimization algorithm to solve (\ref{joint_min}).
Before developing the optimization algorithm, we determine key properties
of (\ref{pej_joint}), (\ref{avg_pe}), and (\ref{joint_min}) that will be utilized.

\begin{property} \label{Proper_invariant}
    The error probability in (\ref{pej_joint}) and therefore the average
    error probability (\ref{avg_pe}) are invariant to the
    scaling of ${w}_j$ by a positive constant.
\end{property}
\noindent Proof is obvious and therefore omitted.
Based on this property, one can set $|w_j|=1$ in the error probability of
(\ref{pej_joint}) and (\ref{avg_pe}) and add $|w_j|=1,~1 \leq j \leq K,$ as
additional constraints to (\ref{joint_min}).

\begin{property} \label{Proper_segment}
    For a fixed ${\bf U}$ that satisfies $\rm{Tr}({\bf U}{\bf U}^H) \le \tau$,  we have
    $\underset{\bf w}{\argmin} P_e = [\underset{w_1}{\argmin} P_{e_1}, \allowbreak
    \ldots, \underset{w_K}{\argmin} P_{e_K}]$. In other words,
    $\mathop{\argmin}\limits_{w_j} P_e = \mathop{\argmin}\limits_{w_j} P_{e_j}$.
\end{property}
\noindent This is obvious since each $P_{e_j}$ depends only on $w_j$ but not on
$w_l,~ l\ne j$. Therefore, minimizing (\ref{avg_pe}) over all $w_j$s
could be partitioned into $K$ decoupled minimizations of $P_{e_j}$ over
$w_j$s, for $1 \le j \le K$.

If one wants to solve (\ref{joint_min}) over ${\bf w}$ for a given ${\bf U}$,
it is clear from Properties \ref{Proper_invariant} and \ref{Proper_segment} and Proposition \ref{lem_positive}
that when there exists no error floor, without loss of
generality, the constraints $|w_j|=1$ and (\ref{eq_posit_constr}), for $1 \le j \le K$, could
be added to the optimization problem. Therefore, for a given ${\bf U}$,
the $K$ optimization problems arising from (\ref{joint_min}) could be rewritten as follows:
\begin{align}\label{pej_min_equal}
    \nonumber
    & \min_{w_j} \frac{1}{N_{b}}
    \sum_{b = 1}^{N_b}
    {Q\left(\frac{\sqrt{2}}{\sigma_z}s_{b,j}{\rm{Re}}\{w_j \sum_{l=1}^K{{\bf h}_j
    {\bf u}_l {s}_{b,l}} \}\right)} \\
    \nonumber
    & {\text{subject to}} \quad |w_j| = 1, \\
    & \quad\quad\quad\quad\quad~ {s_{b,j}{\rm{Re}}\left\{w_j \sum_{l=1}^K{{\bf h}_j
    {\bf u}_l {s}_{b,l}} \right\}} \geq 0,\quad~~ 1 \leq b \leq N_b,
\end{align}
for $1 \le j \le K$. Now, we have the following proposition:
\begin{proposition}\label{Propos_unique_global_joint}
    If the constraints in the minimization problem (\ref{pej_min_equal}) are
    satisfied, any local minimizer of error probability function $P_{e_j}$, i.e.,
    the objective function of the optimization
    problem (\ref{pej_min_equal}), is also a global minimizer. Moreover, the
    global minimizer is unique.
\end{proposition}
\begin{IEEEproof}
    See Appendix \ref{append_uniqe_global_joint}.
\end{IEEEproof}

\begin{property} \label{Proper_convex_joint}
    $\min\limits_{w_j} P_{e_j}$, where $P_{e_j}$ is calculated as in (\ref{pej_joint}),
    could be transformed to a convex optimization problem with a unique global minimizer.
\end{property}
\begin{IEEEproof}
    See Appendix \ref{append_Proper_convex_joint}.
\end{IEEEproof}
Hence, based on Properties \ref{Proper_invariant}-\ref{Proper_convex_joint},
(\ref{joint_min}) is written as
\begin{subequations}\label{joint_min_modified}
    \begin{align}
        &\min_{{\bf w}, {\bf U}} \frac{1}{KN_b} \sum_{j=1}^K
        \sum_{b = 1}^{N_b}
        {Q\left({\frac{\sqrt{2}}{\sigma_z}}s_{b,j}{\rm{Re}}\{w_j \sum_{l=1}^K{{\bf h}_j
        {\bf u}_l {s}_{b,l}} \}\right)}
        \label{joint_min_modified_objective} \\
        &{\text {subject to}}  \quad
        {\rm Tr}({\bf U}{\bf U}^H) \leq \tau,
        \label{joint_min_modified_first_const} \\
        & \quad\quad\quad\quad\quad~
        |w_j| \leq 1, \quad\quad\quad~~\! 1 \leq j \leq K , 
        \label{joint_min_modified_second_const} \\
        &{s_{b,j}{\rm{Re}}\{w_j \sum_{l=1}^K{{\bf h}_j
        {\bf u}_l {s}_{b,l}} \}} \geq 0,~ 1 \leq j \leq K, ~1 \leq b \leq N_b.
    \end{align}
\end{subequations}

\begin{property} \label{Proper_convex_joint_min2}
    Optimization problem (\ref{joint_min_modified}) is a convex optimization problem
    with respect to ${\bf U}$.
\end{property}
\begin{IEEEproof}
    See Appendix \ref{append_convex_joint_min2}.
\end{IEEEproof}

Using Properties \ref{Proper_convex_joint} and \ref{Proper_convex_joint_min2},
we propose the alternating minimization algorithm of Table \ref{table_joint}
for problem (\ref{joint_min}). It should be noted that in Minimization 1 of Table \ref{table_joint},
when (\ref{joint_min_modified_objective}) is minimized over ${\bf U}$, constraints
in (\ref{joint_min_modified_second_const}) do not depend on the optimization
parameter and are therefore not included.
\begin{table}[tp]
    \centering
    \caption{Alternating minimization algorithm for solving (\ref{joint_min})}\label{table_joint}
    \begin{tabular}{l}
        \hline\hline
        \hspace{-.13in}\begin{minipage}[]{3.45in}
            \begin{algorithmic}
                \STATE
                \STATE {\bf Initialization:} \\
                \STATE ${\bf U_0} \gets$ random complex $M \times K$ matrix.\\
                \STATE ${\bf U_0} \gets \frac{{\bf U_0}}{\|{\bf U}_0\|_F} \sqrt{\tau}$.\\
                \STATE $w_0 \gets$ normalized random complex number.\\
                \STATE ${\bf w_0} \gets [w_0]_{1 \times K}$.\\
                \STATE Initialize ${\bf P_e^1} = [P_{e_1}^1,\cdots,P_{e_K}^1]$
                and ${\bf P_e^2} {\bf \prec P_e^1}$ with
                proper values to start the while loop, e.g. ${\bf P_e^1} =
                [1]_{1 \times K}$ and ${\bf P_e^2} = [0.5]_{1 \times K}$.\\
                \WHILE  {$\frac{\sum_{j=1}^K P_{e_j}^1-P_{e_j}^2}{K} > P_e^{\text{threshold}}$}
                \STATE ${\bf P_e^1} \gets {\bf P_e^2}$.\\
                \STATE {\bf \underline{Minimization 1:}}\\
                \vspace{.02 in}
                \STATE Minimize (\ref{joint_min_modified}) over ${\bf U}$ assuming that
                ${\bf w} = {\bf w_0}$ by using a numerical convex optimization method and the initial value
                of ${\bf U_0}$ for ${\bf U}$.\\
                ${\bf U_0} \gets {\bf U}^{\text{opt}}$.\\
                \STATE {\bf \underline{Minimization 2:}}\\
                \FOR { $j=1:K$ }
                \STATE Minimize (\ref{pej_min}) over $w_j$ assuming ${\bf U} = {\bf U_0}$
                by using a numerical convex optimization method and the initial value of
                ${w_0}_j$ for ${w}_j$ .\\
                ${w_0}_j \gets {w}^{\text{opt}}_j$.\\
                \STATE $P_{e_j}^2 \gets P_{e_j}({\bf U_0},{w_0}_j)$.\\
                \ENDFOR
                \ENDWHILE
                \STATE ${\bf U_0}$ and ${\bf w_0}$ are the transmit precoding weights and
                receive filter coefficients, respectively.\\
            \end{algorithmic}
        \end{minipage}
        \\
        \hline\hline
    \end{tabular}
\end{table}
It is easily shown that the algorithm in Table \ref{table_joint} converges to a local
minimum.
\begin{theorem} \label{Theo1}
    The algorithm in Table \ref{table_joint} converges to a local minimum of (\ref{joint_min}).
\end{theorem}
\begin{IEEEproof}
    Since the objective function $P_e({\bf U},{\bf w})$ is minimized at both Minimization
    1 and 2, each iteration causes $P_e$ to be non-increasing. Moreover, considering the fact that
    $P_e$ is always nonnegative, $0 \leq P_e \leq 1$, it guarantees that the algorithm converges
    to a stationary (minimum) point.
\end{IEEEproof}

\section{User Selection}\label{secUserSelection_bpsk}
In this section, we develop a user selection algorithm by taking into
account that MPE transmit precoding is utilized at the transmitter.
It is assumed that in total $K_T$ users are in the system such that $K_T \gg M$ and the
set of all users is given by $\mathcal{A} = \{ 1,\cdots, K_T \}$. For a given time period,
the transmitter selects a subset, $\mathcal{S}$, of
$K$ users out of $K_T$ users, $\mathcal{S} \subseteq \mathcal{A}$, for transmission.
The users should be selected in such a way that certain criteria are met.
We are interested in maximizing the number of selected users and minimizing
the error probabilities at the same time. Since all users receive BPSK
signals, maximizing the number of users may also increase the
throughput. Hence, ideally we would like to solve the
following multiobjective optimization problem which, compared to
(\ref{joint_min_modified}), only has an extra objective function:
\begin{align}\label{multi_obj_user_selection}
    \nonumber
    &\min_{{\bf w}, {\bf U},\mathcal{S} \subseteq \mathcal{A}} \frac{1}{KN_b}
    \sum_{j \in \mathcal{A}} \sum_{b = 1}^{N_b}
    {Q\left({\frac{\sqrt{2}}{\sigma_z}}s_{b,j}{\rm{Re}}\{w_j \sum_{l\in \mathcal{A}}{{\bf h}_j
    {\bf u}_l {s}_{b,l}} \}\right)} \\
    \nonumber
    &\max_{{\bf w}, {\bf U},\mathcal{S} \subseteq \mathcal{A}} K \\
    \nonumber
    &{\text {subject to}} \quad {\rm Tr}({\bf U}{\bf U}^H) \leq \tau,\\
    \nonumber
    &|w_j| \leq 1, \qquad\qquad\qquad\qquad~\, j \in \mathcal{S},\\
    &{s_{b,j}{\rm{Re}}\{w_j \sum_{l \in \mathcal{S}}{{\bf h}_j
    {\bf u}_l {s}_{b,l}} \}} \geq 0,~ j\in \mathcal{S}, ~1 \leq b \leq N_b,
\end{align}
where $K = |\mathcal{S}|$ is the cardinality of the set of selected users.
In the majority of
existing user selection algorithms, e.g. \cite{Goldsmith06,Mao12}, the number of
selected users $K$ should be less than or equal to the degrees of freedom (DoF)
of the system, which in this case is equal to the number of transmit antennas $M$.
However, in our proposed user selection algorithm,
the number of selected users could be more than the DoF of the system,  as will be
confirmed in the numerical results. The increase in the number of selected users
is made possible because in the proposed
user selection algorithm the constraints of the optimization (\ref{multi_obj_user_selection})
originate from the MPE precoding problem which in turn exploits a one-dimensional
discrete modulation rather than a continuous Gaussian input distribution.

Practically, it is more suitable to solve the user selection and precoding problems separately.
Although suboptimal, (\ref{multi_obj_user_selection}) is separated
into a minimization over ${\bf U}$ and ${\bf w}$ and a maximization over
${\mathcal S}$. Since the precoding problem has been addressed in Section \ref{secMpe_bpsk},
user selection will be the only focus in this section. Therefore, instead of
(\ref{multi_obj_user_selection}) we solve
\begin{subequations}\label{user_selection}
    \begin{align}
        &\max_{\mathcal{S} \subseteq \mathcal{A}} |\mathcal{S}| \\
        \label{user_selection_first_const}
        &{\text {subject to}} \quad {\rm Tr}({\bf U}{\bf U}^H) = \sum_{j \in \mathcal{S}}
        \|{\bf u}_j\|^2 \leq \tau,\\
        \label{user_selection_second_const}
        &|w_j| \leq 1, \qquad\qquad\qquad\qquad~~\! j \in \mathcal{S}, \\
        \label{user_selection_third_const}
        &{s_{b,j}{\rm{Re}}\{w_j \sum_{l \in \mathcal{S}}{{\bf h}_j
        {\bf u}_l {s}_{b,l}} \}} \geq 0,~ j\in \mathcal{S}, ~1 \leq b \leq N_b.
    \end{align}
\end{subequations}
In other words, (\ref{user_selection}) maximizes the cardinality of the set of selected users
such that none of the selected users experience error floors and the
transmit power constraint is met, i.e., the number of selected users is maximized subject
to the constraint set of the MPE precoding problem. This means that we try to find the
maximum number of users such that the feasible region of the MPE precoding problem
is not an empty set.
This combinatorial optimization problem has very high
computational complexity. Therefore, we try to find a suboptimal
algorithm for finding a {\it good} set of selected users.

For a tractable solution, we are interested in user selection as a separate entity from
precoding. Therefore, we try to remove the dependence of the optimization problem
(\ref{user_selection}) from the transmit precoding and receive filtering weights.
But before doing this, we first simplify the problem by finding a lower bound for the left
side of the inequality of the constraint (\ref{user_selection_third_const}) as
\begin{equation*}
    {s_{b,j}{\rm{Re}}\{w_j \!\sum_{l \in \mathcal{S}}{{\bf h}_j {\bf u}_l {s}_{b,l}} \}} \! \geq \!
    {\rm{Re}}\{w_j {\bf h}_j {\bf u}_j \} - \!\sum_{\substack{l \in \mathcal{S} \\ l \ne j}}
    |{{\rm{Re}}\{w_j {\bf h}_j {\bf u}_l \}}|.
\end{equation*}
This lower bound is obtained by using the fact $s_{b,j}s_{bl} = \pm 1$ and $x \ge -|x|$.
Since this lower bound is independent of $b$, all $N_b$ constraints for a fixed $j$ in
(\ref{user_selection_third_const}) are replaced with the single constraint:
\begin{equation}\label{mod_constraint}
    {\rm{Re}}\{w_j {\bf h}_j {\bf u}_j \} - \sum_{\substack{l \in \mathcal{S} \\ l \ne j}}
    |{{\rm{Re}}\{w_j {\bf h}_j {\bf u}_l \}}| \ge 0,~  j \in \mathcal{S}.
\end{equation}
Replacing (\ref{user_selection_third_const}) with (\ref{mod_constraint})
reduces the number of constraints in the optimization problem
(\ref{user_selection}). Now, if we further assume that the receivers use ML detection
with normalized filter coefficients we have $w_j = \frac{{\bf u}_j^H{\bf h}_j^H}
{|{\bf u}_j^H{\bf h}_j^H|}$ and therefore
(\ref{user_selection_second_const}) could be removed from (\ref{user_selection}).
The modified constraint of (\ref{user_selection_third_const}), i.e., (\ref{mod_constraint})
is then substituted by the following bound:
\[
    \sum_{\substack{l \in \mathcal{S} \\ l \ne j}}
    |{{\rm{Re}}\{{\bf u}_j^H{\bf h}_j^H {\bf h}_j {\bf u}_l\}}|
    \le
    |{\bf h}_j {\bf u}_j|^2.
\]

Now, to remove the dependence of the user selection problem
on transmit precoding vectors ${\bf u}_j$s,
normalized maximum ratio transmission (conjugate beamforming) is assumed for transmission such that
${\bf u}_j = {\bf h}_j^H \sqrt{\frac{P_T}{\sum_{l \in \mathcal{S}}{|{\bf h}_l|^2}}}$
\cite{Larsson15,Marzetta15}.
Consequently, this removes (\ref{user_selection_first_const}) and
simplifies (\ref{user_selection_third_const}) further to
\[
    \sum_{\substack{l \in \mathcal{S} \\ l \ne j}}
    |{{\rm{Re}}\{{\bf h}_j {\bf h}_l^H\}}|
    \le
    \|{\bf h}_j\|^2.
\]
Therefore, the user selection optimization problem has been simplified to
\begin{subequations}\label{selection_channel}
    \begin{align}
        &\max_{\mathcal{S} \subseteq \mathcal{A}} |\mathcal{S}| \\
        \label{selection_channel_first_const}
        &{\text {subject to}} \quad
        \sum_{\substack{l \in \mathcal{S} \\ l \ne j}}
        |{{\rm{Re}}\{{\bf h}_j {\bf h}_l^H\}}| \le \|{\bf h}_j\|^2, ~ \forall j \in \mathcal{S}.
    \end{align}
\end{subequations}
This problem could be interpreted as follows:
There is an $M$ dimensional vector space on the complex field $\mathds{C}$, i.e., $\mathds{C}^M$,
in which  there are $K_T \gg M$ given elements. We want to choose the maximum number
of elements with the following properties: 
$
    \sum_{\substack{l \in \mathcal{S} \\ l \ne j}}
    |{{\rm{Re}}\{{\bf h}_j {\bf h}_l^H\}}| \le \|{\bf h}_j\|^2,~  \forall j \in
    \mathcal{S}.
$

Now, we try to simplify the problem by reducing the feasible region.
A sufficient condition for (\ref{selection_channel_first_const}) to hold
is that
\[
    |{{\rm{Re}}\{{\bf h}_j {\bf h}_l^H\}}| \le \frac{\min(\|{\bf h}_j\|^2,\|{\bf h}_l\|^2)}
    {|\mathcal{S}|-1}, ~ \forall j,l \in \mathcal{S}.
\]
Substituting (\ref{selection_channel_first_const}) with the above sufficient condition
gives us
\begin{align}\label{selection_chan_simplified}
    \nonumber
    &\max_{\mathcal{S} \subseteq \mathcal{A}} |\mathcal{S}| \\
    &{\text {subject to}} \quad
    \frac{|{{\rm{Re}}\{{\bf h}_j {\bf h}_l^H\}}|}{\min(\|{\bf h}_j\|^2,\|{\bf h}_l\|^2)} \le
    \frac{1}{|\mathcal{S}|-1}, ~ \forall j,l \in \mathcal{S}.
\end{align}

From a geometric point of view, problem (\ref{selection_chan_simplified})
is similar to packing lines
in a Grassmannian manifold of dimension $M$, 
i.e., $G(1,M)$ \cite{Love03,Sloane96,Tse02,Blostein11}.
Namely, problem (\ref{selection_chan_simplified}) is to pack the
$M$-complex space with the maximum number of
lines passing through the origin such that {\it real correlation distance} between any two
lines is less than some value, where we define the real correlation distance between two
lines as
\[
    d_{\text{RC}}({\bf h}_j,{\bf h}_l) \triangleq \frac{|{{\rm{Re}}\{{\bf h}_j {\bf h}_l^H\}}|}
    {\min(\|{\bf h}_j\|^2,\|{\bf h}_l\|^2)}.
\]
It should be remarked that $d_{\text{RC}}$ is not actually a distance or metric but only
possess some properties of a metric.
To solve (\ref{selection_chan_simplified}) we propose the geometric user selection (GUS)
algorithm of Table \ref{table_gus}.
\begin{table}[tp]
    \centering
    \caption{Geometric user selection algorithm}
    \label{table_gus}
    \begin{tabular}{l}
        \hline\hline
        \hspace{-.13in}
        \begin{minipage}[]{3.45in}
            \begin{algorithmic}
                \STATE
                \STATE {\bf Initialization:} \\
                \STATE $iter = 0$.
                \STATE $K^{iter+1} = M+1$.
                \LOOP
                    \STATE $iter \gets iter + 1$.
                    \STATE $i = 1$.
                    \STATE $\mathcal{S}^{iter} = \emptyset$.
                    \STATE $\mathcal{A}_i^{iter} = \{1,\cdots, K_T\}$.
                    \STATE $\mathcal{C}_i^{iter} = \mathcal{A}_i^{iter}$.
                    \STATE {\bf \underline{Main Body of Algorithm:}}\\
                    \WHILE  {$i \le K^{iter}$ and $\mathcal{C}_i \ne \emptyset$}
                        \STATE $\pi_i^{iter} = \operatornamewithlimits{argmax}\limits_{j
                        \in \mathcal{C}_i^{iter}} \|{\bf h}_j\|^2$.\\
                        \STATE $\mathcal{S}^{iter} = \mathcal{S}^{iter} \cup \{\pi_i^{iter}\}$.\\
                        \STATE $\mathcal{A}_i^{iter} \gets \mathcal{A}_i^{iter}\setminus \{\pi_i^{iter}\}$  .
                        \STATE $\mathcal{A}_{i+1}^{iter} = \mathcal{A}_i^{iter}\setminus
                        \{ \forall j \in \mathcal{A}_i^{iter} |
                        d_{\text{RC}}({\bf h}_j,{\bf h}_{\pi_i^{iter}})
                        > \frac{1}{K^{iter}-1}\}$.
                        \STATE $\mathcal{C}_{i+1}^{iter} = \{ \forall j \in \mathcal{A}_{i+1}^{iter}|
                        d_{\text{RC}}({\bf h}_j,{\bf h}_{\pi_i^{iter}})
                        > \frac{\alpha}{K^{iter}-1} \}$. 
                        \IF {$\mathcal{C}_{i+1}^{iter} = \emptyset$}
                            \STATE $\mathcal{C}_{i+1}^{iter} = \mathcal{A}_{i+1}^{iter}$.
                        \ENDIF
                        \STATE $i \gets  i+1$.
                    \ENDWHILE
                    \STATE {\bf \underline{Decision for Next Iteration}}\\
                \ENDLOOP
            \end{algorithmic}
        \end{minipage}
        \\
        \hline\hline
    \end{tabular}
\end{table}

This suboptimum low-complexity algorithm semi-greedily\footnote{The algorithm is semi-greedy
rather than greedy since it does not select the best channel at each iteration. Nonetheless,
it first removes some of the users which are not orthogonal to the last selected users and then
selects the best user among the remaining ones.} selects users
based on both their channel strength and their real correlation distance.
In the Main Body of Algorithm in Table \ref{table_gus}, first
the user with the strongest channel in the set of candidate users, $\mathcal{C}$, is opted
for the set of selected users $\mathcal{S}$.
Then, any user from the set of available users, $\mathcal{A}$, with real correlation
distance of more than $\frac{1}{K-1}$ from the previously selected user, $\pi_i$, is removed
for the next set of available users.
To make the packing tighter, the algorithm selects users from a set of candidate users
rather than from all available users. A user is considered for the next set of candidate users if
its distance from the last selected user, $\pi_i$, is close to the upper
bound $\frac{1}{K-1}$. The parameter $\alpha$ determines how close the distance of the next
candidate users from the last selected user should be to the upper bound.
The Main Body of Algorithm iterates until either $K$ users are
selected or the set of candidate users is empty.
It should be remarked that the cardinality of the selected set in this algorithm is upper bounded
by the initial guess for the number of users $K$.

In Decision for Next Iteration in Table \ref{table_gus}, it is decided whether the set of selected
users could be improved either in the sense of size or the distance among users.
If $iter=1$, then depending on whether the cardinality of
the selected set is equal to $K^{iter}$ or less than $K^{iter}$, $K$ is incremented or decremented
respectively, i.e., $K^{iter+1} = K^{iter} + 1$ or $K^{iter+1} = K^{iter} - 1$.
As a result the number of iterations
is always greater than one. If $iter \ne 1$ then one of four different scenarios may occur:
\begin{enumerate}
  \item If $K^{iter} > K^{iter-1}$ and the size of the current selected set
  is larger than or equal to the size of the previous selected set (
  $|\mathcal{S}^{iter}| \ge |\mathcal{S}^{iter-1}|$), then
  the algorithm gets greedy and prepares to check if the size of the selected set could be further improved
  by incrementing $K$ for the next iteration, i.e., $K^{iter+1} = K^{iter} + 1$.
  \item  If $K^{iter} > K^{iter-1}$ and $|\mathcal{S}^{iter}| < |\mathcal{S}^{iter-1}|$
  then the previous selected set is considered as the selected set ($\mathcal{S} = \mathcal{S}^{iter-1}$)
  and the algorithm breaks from the loop.
  \item  If $K^{iter} < K^{iter-1}$ and $|\mathcal{S}^{iter}| \ge K^{iter}-1$
  then the current selected set is chosen as the
  selected set ($\mathcal{S} = \mathcal{S}^{iter}$) and the algorithm breaks from the loop.
  \item  Otherwise, $K^{iter+1} = K^{iter}-1$.
\end{enumerate}

\begin{remark}
    Now, we justify the choice of $K=M+1$ as the initial value for $K$ in
    GUS algorithm.
    In the context of packing, geodesic distance and chordal distance are
    common metrics with existing bounds for packing problems.
    Therefore, to find an approximate value for the maximum number
    of users $K$, the size of
    the feasible region in (\ref{selection_chan_simplified}) is reduced, again by
    replacing the constraint of (\ref{selection_chan_simplified}) with a sufficient condition as
    \begin{subequations}\label{selection_chan_real_removed}
        \begin{align}
            &\max_{\mathcal{S} \subseteq \mathcal{A}} |\mathcal{S}| \\
            &{\text {subject to}} &&
            \frac{|{\bf h}_j{\bf h}_l^H|}{\min{(\|{\bf h}_j\|^2,\|{\bf h}_l\|^2)}}
            \le \frac{1}{|\mathcal{S}|-1},  ~ \forall j,l \in \mathcal{S},
            \label{selection_chan_real_removed_const}
        \end{align}
    \end{subequations}
    allowing us to utilize existing bounds from differential geometry.

    Next, we find an approximation for the maximum number of lines that could
    fill the $M$ space such that (\ref{selection_chan_real_removed_const}) is satisfied.
    In $G(1,M)$, the principal angle between lines ${\bf h}_j$ and ${\bf h}_l$ is
    defined as \cite{Barg02}
    \begin{equation*}
        \theta_{j,l} = \arccos{\frac{|\langle{\bf h}_j,{\bf h}_l\rangle|}{\|{\bf h}_j\|\|{\bf h}_l\|}}
    \end{equation*}
    and the chordal distance between two lines is defined as \cite{Love03,Sloane96}
    \begin{equation}\label{chordal_distance}
        d_c({\bf h}_j,{\bf h}_l) = |\sin(\theta_{j,l})| = \sqrt{1- \frac{|{\bf h}_j{\bf h}_l^H|^2}
        {\|{\bf h}_j\|^2\|{\bf h}_l\|^2}}.
    \end{equation}
    Similar to the Rankin bound for spherical codes,
    there exists the following upper bound on the chordal distance
    between any two lines in the Grassmannian manifolds for packing $K$ lines
    \cite{Love03,Sloane96}:
    \begin{equation}\label{rankin_bound}
        d_c^2 \le \left\{
        \begin{array}{cc}
          \frac{(M-1)K}{M(K-1)} & \quad\text{if } K \le \binom{M+1}{2}  \\
          \frac{(M-1)}{M} & \quad\text{if } K > \binom{M+1}{2}.
        \end{array}
        \right.
    \end{equation}
    Using (\ref{chordal_distance}) and (\ref{rankin_bound}) we have
    \begin{equation*}
        \left\{
        \begin{array}{cc}
            \!\!\!\!\frac{|{\bf h}_j{\bf h}_l^H|^2}{\min{(\|{\bf h}_j\|^4,\|{\bf h}_l\|^4)}} \!\ge\!
            \frac{|{\bf h}_j{\bf h}_l^H|^2}{\|{\bf h}_j\|^2\|{\bf h}_l\|^2}
            \!\ge\! 1\!-\! \frac{(M-1)K}{M(K-1)} & \text{if } K \!\le\! \binom{M+1}{2} \\
            \!\!\!\!\frac{|{\bf h}_j{\bf h}_l^H|^2}{\min{(\|{\bf h}_j\|^4,\|{\bf h}_l\|^4)}} \!\ge\!
            \frac{|{\bf h}_j{\bf h}_l^H|^2}{\|{\bf h}_j\|^2\|{\bf h}_l\|^2}
            \!\ge\! 1\!-\! \frac{(M-1)}{M} & \text{if } K \!>\! \binom{M+1}{2}
        \end{array}
        \right.
    \end{equation*}
    and by taking (\ref{selection_chan_real_removed_const}) into account, we have
    \begin{equation*}
        \left\{
        \begin{array}{cc}
            1 - \frac{(M-1)K}{M(K-1)} \le \frac{1}{(K-1)^2} &
            ~\text{if } K \le \binom{M+1}{2}\\
            1- \frac{(M-1)}{M} \le \frac{1}{(K-1)^2} &
            ~\text{if } K > \binom{M+1}{2} .
        \end{array}
        \right.
    \end{equation*}
    This is equivalent to the following sufficient (but not necessary) condition:
    \begin{equation*}
        K \le M+1,
    \end{equation*}
    which serves as an approximation to the number of selected users for the first iteration
    in the GUS algorithm.
\end{remark}

\subsection{Complexity Analysis}
Similar to \cite{Evans06}, we analyze the computational complexity of the proposed
GUS algorithm, using flop counts by assuming that a real addition, multiplication,
division, or comparison is counted as one flop. The computational complexity of the
GUS algorithm is calculated as follows:
\begin{itemize}
    \item It takes $2M$ real multiplications and $2M-1$ real addition
    to calculate $\|{\bf h}_j\|^2$. Therefore, $K_T(4M-1)$ flops are needed, to
    calculate all channel norms. Calculation of the channel norms is only performed
    once at the beginning of the GUS algorithm.
    \item Inside the while loop, it takes $|{\mathcal C}_i^{iter}|$ comparisons
    to find $\pi_i^{iter}$.
    \item To calculate $d_{\text{RC}}({\bf h}_j,{\bf h}_{\pi_i^{iter}})$,
    $4M+2$ flops are required, since channel norms are already calculated.
    Therefore, to construct ${\mathcal A}_{i+1}^{iter}$,
    $(4M+3)|\mathcal{A}_{i}^{iter}|$ flops are required.
    \item Since the real correlation distances are already calculated, it takes
    $|\mathcal{A}_{i+1}^{iter}|$ flops for comparisons to construct $\mathcal{C}_{i+1}^{iter+1}$.
    \item The total number of flops in each while loop is therefore counted
    as $|\mathcal{C}_{i}^{iter}| + |\mathcal{A}_{i}^{iter}|(4M+3) +
    |\mathcal{A}_{i+1}^{iter}|$. At each iteration of the while loop, it is obvious
    that $|\mathcal{A}_{i+1}^{iter}| \le |\mathcal{A}_{i}^{iter}|$. Moreover,
    $|\mathcal{A}_{i}^{iter}| \le K_T $ and $|\mathcal{C}_{i}^{iter}| \le K_T$.
    Therefore, $K_T(4M+5)$ is an upper bound on the number of flops in each while loop.
    \item The maximum number of iterations in the while loop is $K^{iter}$ which
    is upper bounded by $2M$. The maximum number of outer loop iteration is $M$.
    Hence, the $2M^2(4M+5)K_T+(4M-1)K_T$ is an upper bound on the total number of flops $\psi$,
    which indicates that $\psi \thickapprox O(K_T M^3)$.
\end{itemize}
Therefore, the complexity order is the same as that of SUS algorithm \cite{Goldsmith06,Mao12}.

\section{Numerical Results}\label{secResults_bpsk}
\subsection{MPE Precoding Methods}
In this section, we first consider a broadcast channel with a 3-antenna transmitter sending
information 
to 3 users each equipped with 1 antenna.
It is assume that the transmitter has one information symbol for each receiver in the
same time and frequency slots.
The channel gains are assumed to be quasi static and follow a
Rayleigh distribution. Since our focus is on various precoding methods rather
than on the effects of channel estimation, we assume that perfect CSI of all channels is
available at the transmitter and
each receiver only has perfect knowledge of its own channels \cite{Peel05,Sadek07WC}.
At the receivers, white Gaussian noise is added to the received signal.

\begin{figure}[tp]
  \centering
  \includegraphics[width=3.49in]{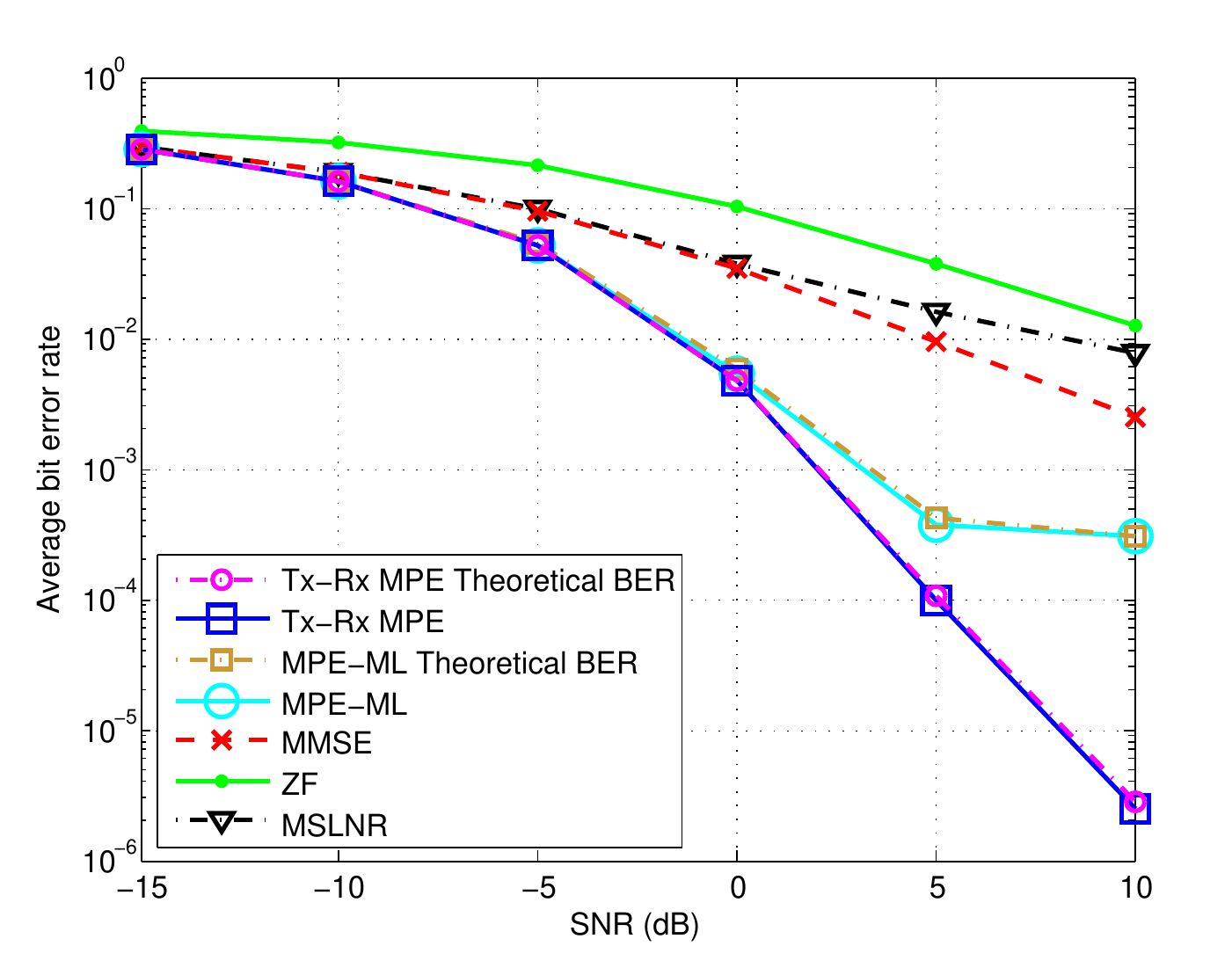}\\ 
  \caption{Bit error rate of users using MSLNR, ZF, MMSE, MPE-ML, and MPE joint
  transmit precoding-receive filtering (Tx-Rx)
  for $M=3$ antenna transmitter and $K=3$ users with BPSK modulation
  and without user selection.}
  \label{fig_ber_no_selection}
\end{figure}

Fig. \ref{fig_ber_no_selection} compares the bit error rates of MSLNR, ZF, MMSE, MPE-ML
(from Section \ref{subSecMpeML_bpsk}), and MPE joint transmit precoding-receive
filtering (Tx-Rx) (from Section
\ref{subSecMpeJoint_bpsk}) methods. For all of these methods, maximum
likelihood is used at the receiver except for MPE joint Tx-Rx in which each
receive filter coefficient is calculated by the transmitter jointly with the transmit
precoding weights and is provided to the corresponding receiver.
As can be seen, MPE transmit precoding methods substantially improve the performance
of all users. For example, at BER of $10^{-2}$ both MPE precoding methods show
a gain of about 6.5dB compared with MMSE precoding and much more gain compared with ZF and
MSLNR.
It should be mentioned that in Fig. \ref{fig_ber_no_selection},
theoretical BER curves of MPE-ML and MPE Tx-Rx methods are obtained by substituting
the calculated precoding weights for each channel realization into the error probability
expressions of (\ref{pe_ML_ub}) and (\ref{joint_min_modified_objective}), respectively.
All other curves in Fig. \ref{fig_ber_no_selection} are the result of Monte Carlo
simulations. As can be seen in Fig. \ref{fig_ber_no_selection} the simulations
confirm the theoretical results.
It is interesting to observe that the performance of MPE-ML precoding closely follows that of the MPE joint
Tx-Rx up to a certain SNR. At higher SNRs, although still outperforming classical
precoding methods, MPE-ML precoding shows a large gap in performance compared with MPE Tx-Rx.
The reason that MPE-ML precoding cannot follow the performance
of MPE Tx-Rx is that in Minimization 1 of Table \ref{table_ML} the
optimization problem is not convex. Therefore, the solver of the optimization
might get stuck at a local minimizer instead of the global minimizer of (\ref{ML_min_mod1}).

\begin{figure}[tp]
  \centering
  \includegraphics[width=3.49in]{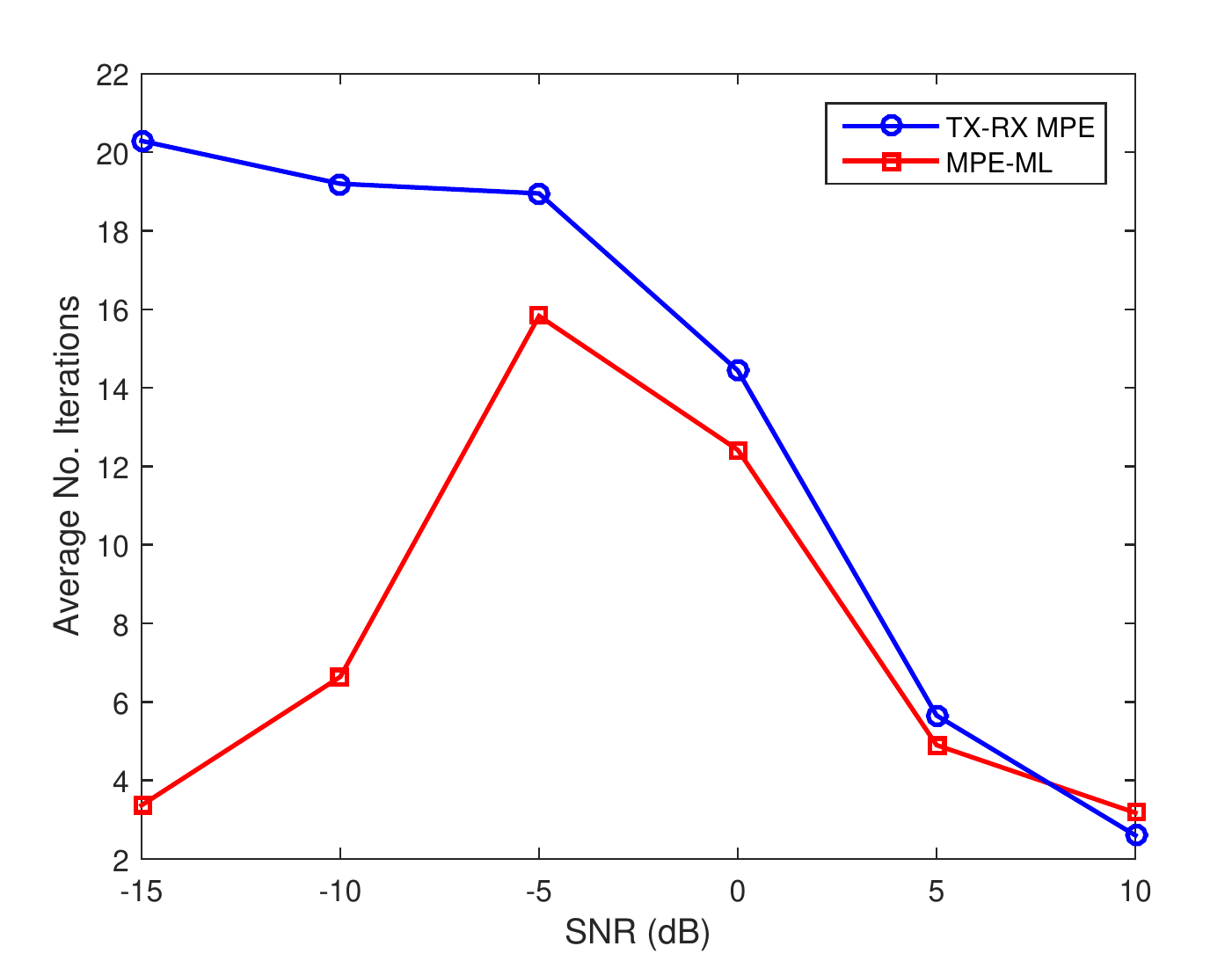}\\ 
  \caption{Average number of iterations for convergence of algorithms of
  Table \ref{table_ML} and \ref{table_joint} when $M=3$ antennas and
  $K=3$ users without user selection.}
  \label{fig_iteration}
\end{figure}

In Fig. \ref{fig_iteration} we show the average number of iterations needed for the
convergence of the algorithms in Tables \ref{table_ML} and \ref{table_joint}.
In both algorithms the average number of iterations is less than 20. It is very
interesting to observe that as SNR increases the number of iterations
needed for the convergence of the algorithm in Table \ref{table_joint} decreases. This
is because $P_e^{\text{threshold}}$ is set fixed at
$10^{-8}$ for all SNRs while the error probability decreases from about
$10^{-1}$ to $10^{-5}$. If it is assumed that the error probability is known in advance,
it would be better to change $P_e^{\text{threshold}}$ to a small fraction
of the error probability to have a faster convergence with the same reliability.

\subsection{User Selection}
\begin{figure}[tp]
  \centering
  \includegraphics[width=3.49in]{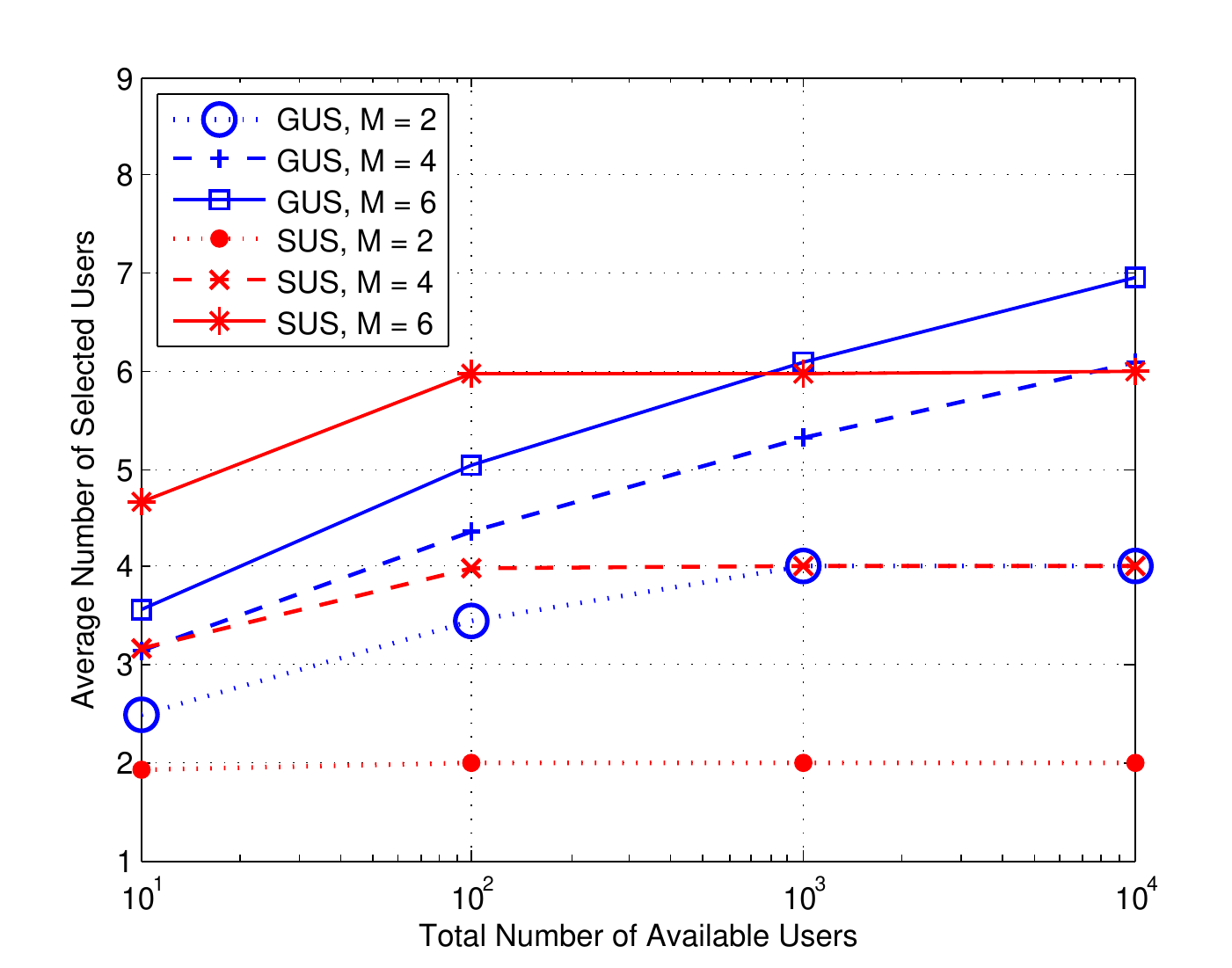}\\ 
  \caption{Average number of selected users vs. total number of available users for
  $M=2,4,6$ antennas.}
  \label{fig_GUS_SUS}
\end{figure}

First, the proposed GUS algorithm is compared with semi-orthogonal user selection (SUS)
algorithm of
\cite{Goldsmith06}. Fig. \ref{fig_GUS_SUS} shows the average number of selected users
over 1,000 different channel realizations versus
the total number of available users for three different cases: when the number of transmit
antennas is 2, 4, and 6. We observe that for SUS it is possible to have
at most as many users as the
number of transmit antennas, as it is obvious by its algorithm in \cite{Goldsmith06},
while GUS may select more users. For example, when there are
$K_T = 10,000$ users available{\footnote{Of course it is not practical to service
10,000 users with one transmitter. This large number of users is just for
illustration purposes to gain insight into the system.}},
GUS can select 6 users with only 4 antennas compared to the 4 selected
users of the SUS algorithm. When the transmitter has 2 antennas
and there are only 10 users available to select from, GUS can select 2.48 users on average which is
still more than the number of
transmit antennas, while SUS selects 1.92 users on average. It should be remarked that in all
our simulations the parameter $\alpha$ in GUS is set to be $\frac{K^{iter}-1}{K^{iter}}$.

As expected, it can also be seen that until saturation, the number of selected users for both
GUS and SUS increases as the total number of available users increases. Each user
corresponds to a line passing through the origin of the complex $M$ dimensional
hypersphere. Hence, when the number of users increases, the density of possible lines to
choose from increases.
This makes it more likely to pack the space with more lines obeying the specific
distance.
It should be remarked that when $M=6$, for smaller numbers of users SUS outperforms
GUS. This indicates the requirement for the development of better algorithms than GUS,
which would not only be able to select more users than the number of transmit antennas as GUS does,
but would also be capable of outperforming SUS in all scenarios.

\begin{figure}[tp]
  \centering
  \includegraphics[width=3.49in]{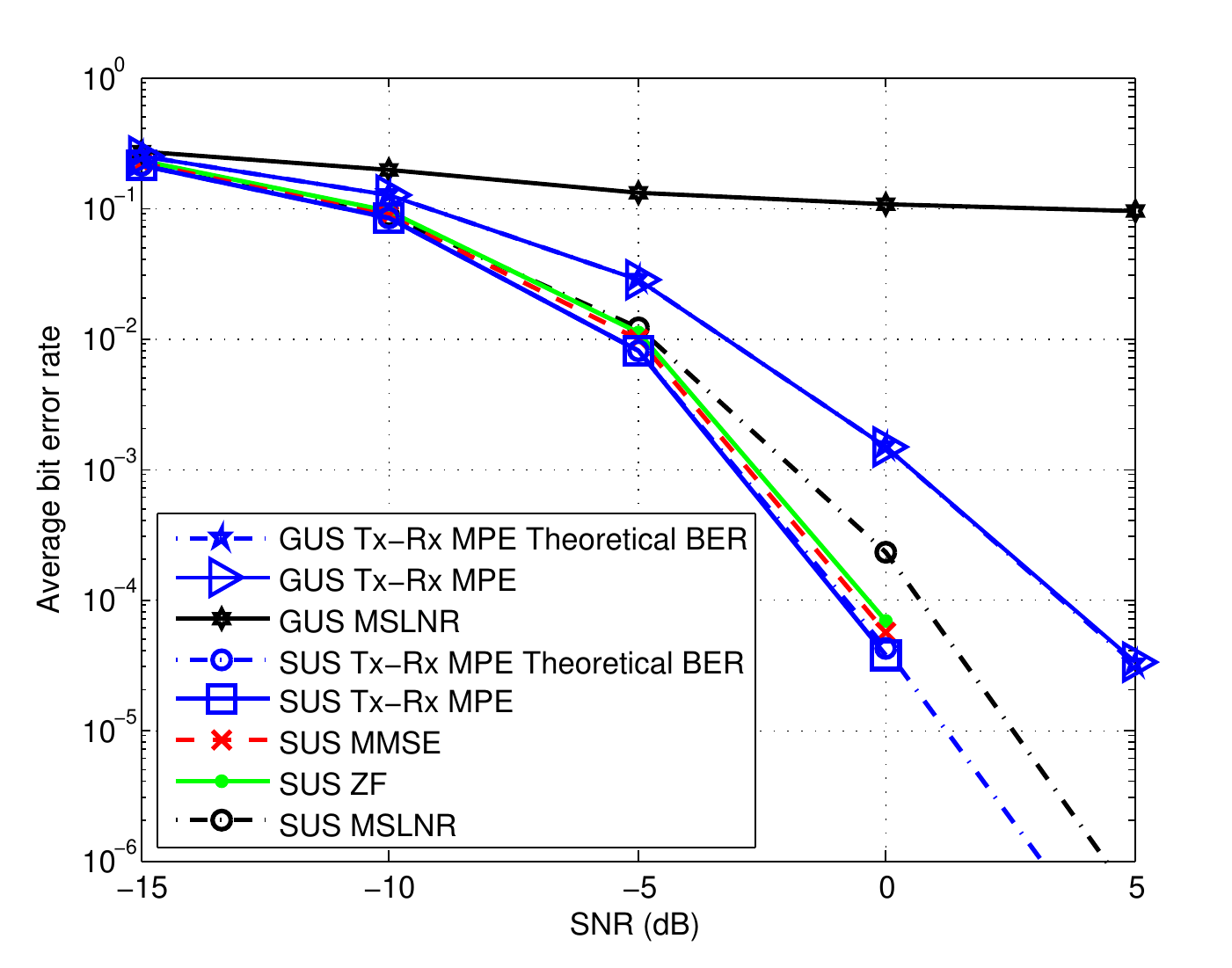}\\ 
  \caption{Average bit error rate of users when $M=2$ antennas
  and $K_T = 50$ users.  In this scenario, the average number of selected users over
  all SNRs is 2 for SUS while it is 3.19 for GUS.}
  \label{fig_Pe_2antennas_GUS_SUS}
\end{figure}

Now, we study the effect of user selection in conjunction with precoding.
Consider a system with a 2-antenna transmitter that first selects a set of users out
of $K_T = 50$ users by using either GUS or SUS, and then sends information to the selected
users using various precoding methods. Fig. \ref{fig_Pe_2antennas_GUS_SUS} shows the error
probabilities of different combinations of user selection and precoding methods.
As can be seen, the lowest error probabilities are achieved by SUS rather than GUS,
and by a small margin, MPE precoding outperforms all other precoding methods used
for SUS. However, it could not be simply concluded that the SUS algorithm outperforms GUS since
the average number of selected users over all SNRs is 2 for SUS while
it is 3.19 for GUS.
In other words, although GUS selects more users, the users have higher
error probabilities in comparison with the selected users of SUS, as expected.
It should be noticed that for the proposed geometric user selection algorithm,
ZF and MMSE precoding performance curves are absent
from Fig. \ref{fig_Pe_2antennas_GUS_SUS}, since the number of selected users by GUS
is larger than the number of transmit antennas and is
therefore not suitable for ZF and MMSE precoding.

\begin{figure}[tp]
  \centering
  \includegraphics[width=3.49in]{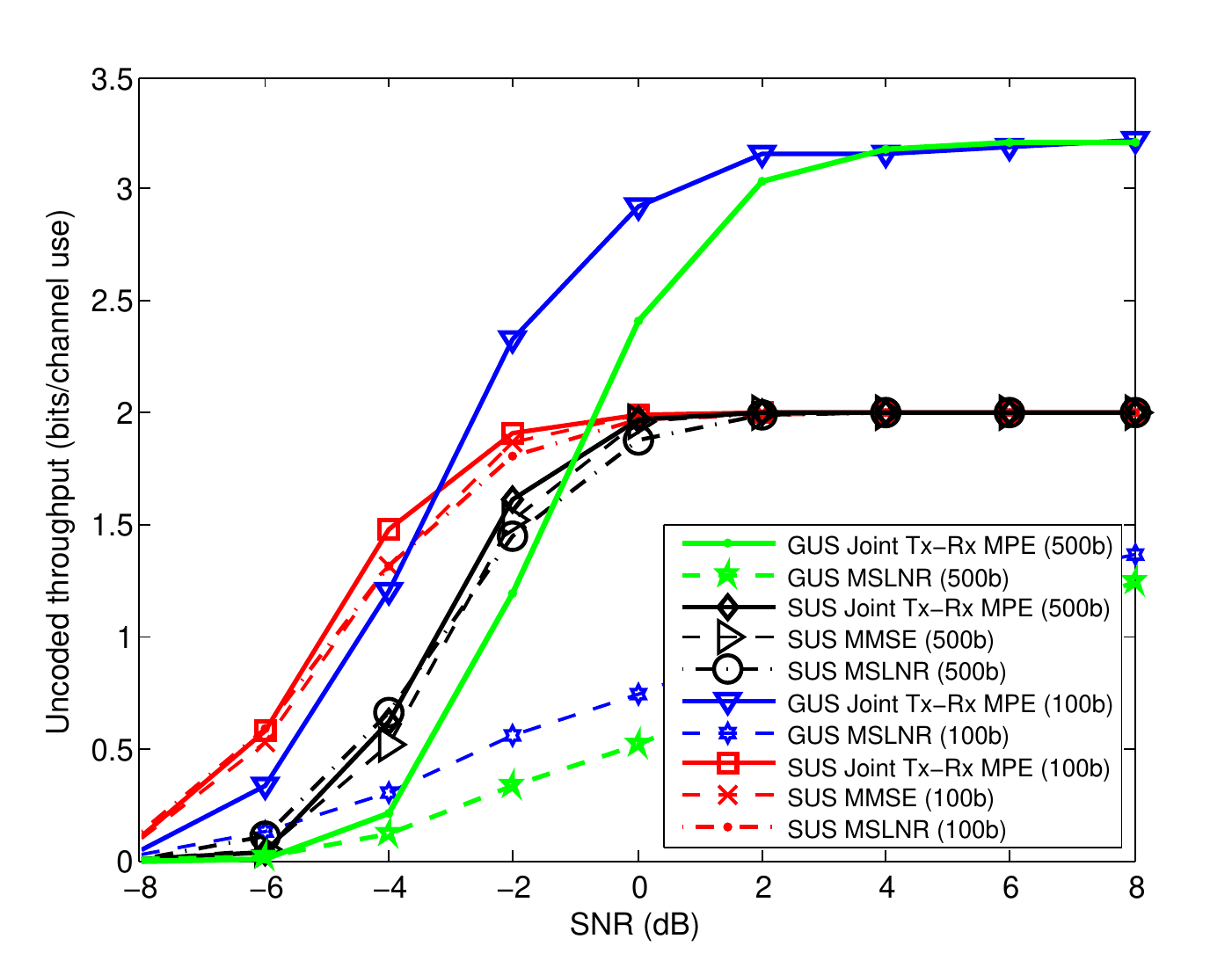}\\ 
  \caption{Average frame-wise rate when $M=2$, $K_T = 50$, and frame sizes of
  100 and 500 bits.}
  \label{fig_rate}
\end{figure}

As observed in Fig. \ref{fig_Pe_2antennas_GUS_SUS}, since the error probability
alone is not a good indicator of the performance
when there are different numbers of users in the system, Fig. \ref{fig_rate}
is provided to give more insight into
the performances of GUS and SUS.
In Fig. \ref{fig_rate} the throughput is shown for different combinations of user
selection and precoding. Similar to \cite{Saeed14}, we use the notion of
expected throughput for frame-based transmission as $E[{\rm{Thr}}] =
(1-P_e)^{\ell} K_{\text{avg}}$ bits per channel use, where $\ell$ is the frame
size and $K_{\text{avg}}$ is the average number of selected users.
We consider two different frame sizes: 100 and 500 bits.
No channel coding is assumed, and a transmission is considered to be successful if
the entire frame is decoded error-free. It can be seen in Fig. \ref{fig_rate} that
as SNR increases the achievable expected throughput approaches limits
determined by the average numbers of selected users by SUS and GUS. For this example,
at higher SNRs the achievable throughput by GUS and MPE precoding is about
$160\%$ of the achievable throughput by SUS and any of the other precoding methods.
Moreover, Fig. \ref{fig_Pe_2antennas_GUS_SUS} shows that as the frame size increases
the throughput decreases. However, for all frame sizes the throughput eventually
approaches its upper limit, which is dictated by the average number of selected users.


\section{Conclusions}\label{secConclusion_bpsk}
In this paper, it has been demonstrated that by exploiting the
modulation type, a transmit precoder can be designed with significantly
improved performance. For BPSK, this amount to minimizing
the average error probability of multiple users. Two algorithms were developed
based on the concepts of alternating minimization and convex optimization for
two different cases: (i) when the receivers use single-user maximum likelihood
detection, (ii) when the transmitter optimizes the transmit precoding weights
and the receive filter coefficients jointly. For BPSK signalling,
it has been shown that minimum probability
of error precoding outperforms conventional transmit precoding methods
such as ZF, MMSE, and MSLNR.
Unfortunately, the results reported here for BPSK do not easily generalize
to enable the combining of transmit precoding with other types of modulation
and instead represents a topic of ongoing and future research.
A very significant finding reported here is the proposed geometric user
selection (GUS) algorithm has been shown to be able to select more users
than the number of transmit antennas as well as increase total
throughput.

\appendices
\section{Proof of Proposition \ref{lem_positive}}\label{append_lem_positive}
Assume that there exists a $b=b_1$ such that for a given $w_j$ and ${\bf U}$,
$s_{b_1,j} {\rm{Re}} \{ w_j \allowbreak \sum_{l=1}^K {{\bf h}_j
{\bf u}_l {s}_{b_1,l}} \} < 0$.
Therefore,
\[\frac{s_{b_1,j}{\rm{Re}}\{w_j \sum_{l=1}^K{{\bf h}_j
    {\bf u}_l {s}_{b_1,l}} \}} {\frac{\sigma_z}{\sqrt{2}} |w_j|} < 0.
\]
Hence,
\begin{equation}\label{eq_positive_proof}
{Q\left(\frac{s_{b_1,j}{\rm{Re}}\{w_j \sum_{l=1}^K{{\bf h}_j
{\bf u}_l {s}_{b_1,l}} \}} {\frac{\sigma_z}{\sqrt{2}} |w_j|}\right)}
> \frac{1}{2}.
\end{equation}
Moreover, if there exists a $b_1$ such that (\ref{eq_positive_proof}) is
true, then there also exists a $\bar b_1$ where each bit in ${\bf s}_{b_1}$ is
inverted. For $b=\bar b_1$ we also have
${Q\left(\frac{s_{\bar b_1,j}{\rm{Re}}\{w_j \sum_{l=1}^K{{\bf h}_j
{\bf u}_l {s}_{\bar b_1,l}} \}} {\frac{\sigma_z}{\sqrt{2}} |w_j|}\right)}
> \frac{1}{2}$,
and therefore,
\begin{equation*}
    P_{e_j}
    > \frac{1}{N_{b}} + \frac{1}{N_{b}} \!
    \sum^{N_b}_{\substack{b = 1\\
    b \ne b_1, \bar b_1}}
    {\!\!\!Q\!\left(\frac{s_{b,j}{\rm{Re}}\{w_j \sum_{l=1}^K{{\bf h}_j
    {\bf u}_l {s}_{b,l}} \}} {\frac{\sigma_z}{\sqrt{2}} |w_j|}\right)}.
\end{equation*}
In other words, there always exists an error floor of $1/N_b$.

\section{Proof of Property \ref{Proper_convex_ML}}\label{append_Proper_convex_ML}
Assuming $\alpha \in \mathds{R}$, we define ${\bf a}_0$ as
\[
{\bf a}_0 = \alpha {\bf a}_1+(1-\alpha){\bf a}_2.
\]
We also define the function $g({\bf a})$ to denote the argument of the $Q$-function
in (\ref{pe_ML_ub}). Therefore,
\begin{align*}
    g({\bf a}_0)  &=
    \frac{\|{\bf h}_j\|{a_0}_j }
    {\frac{\sigma_z}{\sqrt{2}}} +
    \frac{\sum_{\substack{l=1 \\ l \ne j}}^K{{\rm Re}\{{\bf \bar u}_j^H {\bf h}_j^H {\bf h}_j
    {\bf \bar u}_l {s}_{b,l}s_{b,j} ({a_0}_l) \}}}
    {\frac{\sigma_z}{\sqrt{2}}\|{\bf h}_j\|}  \\
    &=\alpha g({\bf a}_1) + (1-\alpha) g({\bf a}_2),
\end{align*}
which means that the argument of each $Q$-function in (\ref{pe_ML_ub})
is affine with respect to ${\bf a}$.
Now, considering the fact that $Q$-function is convex over positive arguments
and that affine mapping does not change the convexity it becomes clear that
$Q(g({\bf a}))$ is a convex function \cite{BoydBook04}.
Taking into account that the sum of two convex functions is a convex
function it could be inferred that (\ref{pe_ML_ub}), i.e., the objective function
of (\ref{ML_min_mod1}), is convex with respect to ${\bf a}$.

It is obvious that the constraint (\ref{ML_min_mod1_first_const}) is a convex set
with respect to ${\bf a}$ since it represents the volume inside of a $K$-dimensional
sphere.
By using the definition of a convex set \cite{BoydBook04}, it could be shown that the constraints
in (\ref{ML_min_mod1_second_const}) are also convex sets with respect to
${\bf a}$. Also as mentioned earlier, the constraints defined in
(\ref{ML_min_mod1_third_const}) are not included in Minimization 2.
Therefore, Minimization 2 of Table \ref{table_ML} is a convex optimization problem with
a convex objective function and closed convex constraints.

\section{Proof of Proposition \ref{Propos_unique_global_joint}}\label{append_uniqe_global_joint}
The minimization problem is considered over the following feasible set:
\begin{equation}\label{constraint_set}
    \mathcal{F}_j\!=\!  \{{w}_j\!:\!|{w}_j| \!=\! 1,
    {s_{b,j}{\rm{Re}}\!\{w_j\! \sum_{l=1}^K{{\bf h}_j
    {\bf u}_l {s}_{b,l}}\! \}} \!\geq \!0, 1 \!\leq\! b \!\leq\!
    N_b\}.
\end{equation}
Assume that ${w}_{j_1} \in \mathcal{F}_j$ is a global minimizer of the
optimization problem (\ref{pej_min_equal}), and ${w}_{j_2} \in \mathcal{F}_j$ is
a local minimizer of the problem such that
\begin{equation}\label{pej_assumption}
    P_{e_j}({ w}_{j_1}) < P_{e_j}({w}_{j_2}) .
\end{equation}
Assuming $0 < \alpha < 1$, we define ${w}_{j_0}$ as
\begin{equation*}
    {w}_{j_0} = \frac {\alpha {w}_{j_1}+(1-\alpha){
    w}_{j_2}} {\| \alpha {w}_{j_1}+(1-\alpha){w}_{j_2} \|}.
\end{equation*}
Therefore, we have $\|{w}_{j_0}\|=1$, and for $1 \leq b \leq
N_b$, we have ${s_{b,j}{\rm{Re}}\left\{w_{j_0} \sum_{l=1}^K{{\bf h}_j
{\bf u}_l {s}_{b,l}} \right\}} \geq 0$. Hence, it can be inferred that ${
w}_{j_0} \in \mathcal{F}_j$. It is also obvious that
\begin{equation*}
    \| \alpha {w}_{j_1}+(1-\alpha){ w}_{j_2} \| \leq \alpha
    \|{w}_{j_1}\|+(1-\alpha)\|{w}_{j_2}\| = 1.
\end{equation*}
Consequently,
\begin{align}\label{arg_inequal}
    \nonumber
    &s_{b,j}{\rm{Re}}\left\{w_{j_0} \sum_{l=1}^K{{\bf h}_j
    {\bf u}_l {s}_{b,l}} \right\} \ge
    \alpha s_{b,j}{\rm{Re}}\left\{w_{j_1} \sum_{l=1}^K{{\bf h}_j
    {\bf u}_l {s}_{b,l}} \right\} + \\
    &(1-\alpha) s_{b,j}{\rm{Re}}\left\{w_{j_2} \sum_{l=1}^K{{\bf h}_j
    {\bf u}_l {s}_{b,l}} \right\},
\end{align}
for $1 \leq b \leq N_b$.
Therefore,
\begin{align}\label{Q_inequal}
    \nonumber
    &{Q\left(\frac{\sqrt{2}}{\sigma_z}s_{b,j}{\rm{Re}}\{w_{j_0}
    \sum_{l=1}^K{{\bf h}_j {\bf u}_l {s}_{b,l}} \}\right)}
    \leq \\
    \nonumber
    &{Q (\! \frac{\alpha s_{b,j}{\rm{Re}}\{\!w_{j_1}\!
    \sum\limits_{l=1}^K{\!{\bf h}_j {\bf u}_l {s}_{b,l}} \!\}
    \!+\!
    (1-\alpha) s_{b,j}{\rm{Re}}\{\!w_{j_2}
    \!\sum\limits_{l=1}^K{\!{\bf h}_j {\bf u}_l {s}_{b,l}}\!\} }
    {\sqrt{\frac{\sigma_z^2}{2}}} \!)} \\
    \nonumber
    & \leq
    \alpha{Q\left(\frac{\sqrt{2}}{\sigma_z}s_{b,j}{\rm{Re}}\{w_{j_1}
    \sum_{l=1}^K{{\bf h}_j {\bf u}_l {s}_{b,l}} \}\right)}
    +\\
    &(1-\alpha){Q\left(\frac{\sqrt{2}}{\sigma_z}s_{b,j}{\rm{Re}}\{w_{j_2}
    \sum_{l=1}^K{{\bf h}_j {\bf u}_l {s}_{b,l}} \}\right)}
\end{align}
where the first inequality is the result of (\ref{arg_inequal}) and due
to the fact that ${\rm Q}(x)$ is a decreasing function for $x \geq
0$, and the second inequality stands because ${\rm Q}(x)$ is a
convex function for $x \geq 0$.

From (\ref{pej_joint}) and (\ref{Q_inequal}), it can be inferred that
\begin{align*}
    \nonumber
    &P_{e_j}({w}_{j_0}) = \frac{1}{N_{b}} \sum_{b = 1}^{N_b}
    {Q\left(\frac{\sqrt{2}}{\sigma_z|w_{j_0}|}s_{b,j}{\rm{Re}}\{w_{j_0}
    \sum_{l=1}^K{{\bf h}_j{\bf u}_l {s}_{b,l}} \}\right)} \\
    \nonumber
    &\le \frac{\alpha}{N_{b}} \sum_{b = 1}^{N_b}
    {Q\left(\frac{\sqrt{2}}{\sigma_z|w_{j_1}|}s_{b,j}{\rm{Re}}\{w_{j_1}
    \sum_{l=1}^K{{\bf h}_j{\bf u}_l {s}_{b,l}} \}\right)} + \\
    \nonumber
    & \frac{1-\alpha}{N_{b}} \sum_{b = 1}^{N_b}
    {Q\left(\frac{\sqrt{2}}{\sigma_z|w_{j_2}|}s_{b,j}{\rm{Re}}\{w_{j_2}
    \sum_{l=1}^K{{\bf h}_j{\bf u}_l {s}_{b,l}} \}\right)} \\
    & = \alpha P_{e_j}({w}_{j_1}) + (1 - \alpha) P_{e_j}({
    w}_{j_2}) < P_{e_j}({w}_{j_2}) , \quad \forall \alpha \in
    (0,1),
\end{align*}
where the last inequality is due to the fact that ${w}_{j_1}$
is the global minimizer of $P_{e_j}({w}_j)$. Now, let $\alpha
\rightarrow 0$, ${w}_{j_0} \rightarrow {
w}_{j_2}$. Hence, in a small neighborhood of ${w}_{j_2}$,
there always exists a ${w}_{j_0}$, so that $P_{e_j}({
w}_{j_0}) < P_{e_j}({w}_{j_2})$, i.e., ${w}_{j_2}$ is not
a local minimizer. In other words, there does not exist any local
minimizer such that (\ref{pej_assumption}) holds. Therefore, it can be
concluded that either no local minimizer exists, which
proves the proposition, or there exists a local minimizer such that
$P_{e_j}({ w}_{j_1}) \geq P_{e_j}({ w}_{j_2})$. However,
since ${w}_{j_1}$ is a global minimizer of $P_{e_j}({
w}_j)$, we have $P_{e_j}({w}_{j_1}) \leq P_{e_j}({
w}_{j_2})$. Therefore, it can be concluded that $P_{e_j}({
w}_{j_1}) = P_{e_j}({w}_{j_2})$, i.e., the local minimizer (if
exists) is also a global minimizer.

To show the uniqueness of the global minimizer, first
the following set is considered:
\begin{equation*}
    \mathcal{F}_j^0 \!=\!  \{ { w}_j : |{w}_j| = 1,
    {\rm{Re}}\{w_j \sum_{l=1}^K{{\bf h}_j
    {\bf u}_l} \} \!=\! 0, ~1 \leq b \leq N_b \}.
\end{equation*}
It is obvious that each point in this set is a global maximizer of
error probability function in (\ref{pej_joint}) constrained by the set
defined in (\ref{constraint_set}), because the arguments in
all $Q$-functions in error probability will be zero. Therefore, to
solve the minimization problem it is sufficient to solve the
problem over the set $\mathcal{F}_j^1=\mathcal{F}_j - \mathcal{F}_j^0$.
$P_{e_j}({w}_j)$ is
strictly convex on $\mathcal{F}_k^1$, because ${\rm Q}(x)$ is strictly
convex for $x > 0$. Assume that ${w}_{j_1} \neq {w}_{j_2}$
are two global minimizers of the optimization problem
(\ref{pej_min_equal}). We define ${ w}_{j_0}$ as follows:
\begin{equation*}
    {w}_{j_0} = \frac {\alpha {w}_{j_1}+(1 - \alpha){
    w}_{j_2}} {\| \alpha {w}_{j_1}+(1 - \alpha){w}_{j_2} \|},
    \quad \forall \alpha \in (0,1).
\end{equation*}
Since ${w}_{j_1}$ is a global minimizer, it is obvious that
\begin{equation}\label{contradiction1}
    P_{e_j}({w}_{j_0}) \geq P_{e_j}({ w}_{j_1}) .
\end{equation}
On the other hand, we have
\begin{equation}\label{contradiction2}
    P_{e_j}({w}_{j_0}) < \alpha P_{e_j}({w}_{j_1}) + (1 -
    \alpha) P_{e_j}({w}_{j_2}) = P_{e_j}({ w}_{j_1}) ,
\end{equation}
because $P_{e_j}({w}_j)$ is strictly convex on $\mathcal{F}_j^1$. Since
(\ref{contradiction2}) contradicts (\ref{contradiction1}), it can be inferred
that the global minimizer is unique.

\section{Proof of Property \ref{Proper_convex_joint}}\label{append_Proper_convex_joint}
Properties \ref{Proper_invariant} and \ref{Proper_segment} and Propositions \ref{lem_positive}, and
\ref{Propos_unique_global_joint} show that without loss of generality
$\min\limits_{w_j} P_{e_j}$ could be stated as (\ref{pej_min_equal})
which does not have any local minimizer and has only one global minimizer.
Problem (\ref{pej_min_equal}) could be equivalently rewritten in the form
of convex optimization problem as
\begin{align}\label{pej_min}
    \nonumber
    & \min_{w_j} \frac{1}{N_{b}}
    \sum_{b = 1}^{N_b}
    {Q\left(\frac{\sqrt{2}}{\sigma_z}s_{b,j}{\rm{Re}}\{w_j \sum_{l=1}^K{{\bf h}_j
    {\bf u}_l {s}_{b,l}} \}\right)} \\
    \nonumber
    & {\text{subject to}} \quad |w_j| \leq 1, \\
    & \quad\quad\quad\quad\quad~ {s_{b,j}{\rm{Re}}\left\{w_j \sum_{l=1}^K{{\bf h}_j
    {\bf u}_l {s}_{b,l}} \right\}} \geq 0,\quad 1 \leq b \leq N_b,
\end{align}
since $|w_j| \le 1$ is an active constraint.

\section{Proof of Property \ref{Proper_convex_joint_min2}}\label{append_convex_joint_min2}
Before stating the proof it should be noted that the ${\rm Re}\{.\}$ operator is not
a linear function.
Assuming $\alpha \in \mathds{R}$,
we define ${\bf U}_0$ as
\[
{\bf U}_0 = \alpha {\bf U}_1+(1-\alpha){\bf U}_2.
\]
We also define the function $g({\bf U})$ to denote the argument of the $Q$-function
in (\ref{joint_min_modified}). Therefore,
\begin{align*}
    g({\bf U}_0)  &= \frac{\sqrt{2}}{\sigma_z}s_{b,j}{\rm{Re}}\{w_j \sum_{l=1}^K{{\bf h}_j
    (\alpha {\bf u}_{1_l}+ (1-\alpha){\bf u}_{2_l}) {s}_{b,l}} \} \\
    &= \alpha g({\bf U}_1) + (1-\alpha) g({\bf U}_2),
\end{align*}
which means that the argument of each $Q$-function in (\ref{joint_min_modified_objective})
is affine with respect to ${\bf U}$.
Now, considering the fact that the $Q$-function is convex for positive arguments
and that affine mapping does not change the convexity,
$Q(g({\bf U}))$ is a convex function \cite{BoydBook04}.
Since the sum of two convex functions is a convex
function it becomes clear that the objective function in (\ref{joint_min_modified}),
$P_e({\bf U})$, is convex with respect to ${\bf U}$.

It is obvious that the constraint (\ref{joint_min_modified_first_const}) is a convex set
with respect to ${\bf U}$ since ${\rm Tr}({\bf U}{\bf U}^H) \le 1$ represents
the interior and boundary of a $KM$-dimensional ball.
It is also interesting to note that ${\rm Tr}({\bf U}{\bf U}^H) = \|{\bf U}\|_F^2$ and
every norm is a convex function.
The second constraint is not defined over ${\bf U}$ and
the third constraint could be shown to be a convex set by using the definition.
Therefore, problem (\ref{joint_min_modified}) is a convex optimization problem
over ${\bf U}$ with
a convex objective function and closed convex constraints.




\end{document}